\documentclass[aps,twocolumn,prd,preprintnumbers, 10pt,superscriptaddress]{revtex4-1}
\pdfoutput=1
\usepackage[hidelinks]{hyperref}
\usepackage{amssymb}
\usepackage{amsfonts}
\usepackage{graphicx}
\usepackage{epstopdf}
\usepackage{dcolumn}
\usepackage{amsmath}
\usepackage{latexsym,bm}
\usepackage{amsthm}
\usepackage{slashed}
\usepackage{float}
\usepackage{color}
\usepackage{url}
\usepackage{longtable}
\usepackage{rotating}
\usepackage[normalem]{ulem}
\usepackage{faktor}
\usepackage{tikz-cd}
\usepackage{tensor}
\usepackage{array}
\numberwithin{equation}{section}

\newcommand{\be}{\begin{equation}}
\newcommand{\ee}{\end{equation}}
\newcommand{\ba}{\begin{aligned}}
\newcommand{\ea}{\end{aligned}}

\definecolor{DarkGreen}{rgb}{0.1, 0.7, 0.3}

\newtheorem{conj}{Conjecture}

\newcommand{\bit}{\begin{itemize}}
\newcommand{\eit}{\end{itemize}}
\newcommand{\ben}{\begin{enumerate}}
\newcommand{\een}{\end{enumerate}}

\newcommand{\Z}{{\mathbb Z}}

\newcommand{\bC}{{\mathbb C}}

\newcommand{\cC}{\mathcal{C}}

\newcommand{\cZ}{\mathcal{Z}}

\newcommand{\C}{\mathsf{C}}

\def\diag{\mathop{\mathrm{diag}}\nolimits}

\def\sign{\mathop{\mathrm{sign}}\nolimits}

\def\unit{{1\kern-.65ex {\rm l}}}
\def\1{{1\kern-.65ex {\rm l}}}

\newcommand{\beq}{\begin{equation}}
\newcommand{\eeq}{\end{equation}}

\usepackage{verbatim}
\usepackage{tikz}
\usetikzlibrary{arrows,shapes.arrows,decorations.markings}
     \tikzset{>=triangle 90}
     \tikzstyle{gr}=[draw,circle,green!50!black,fill=green!50!black,scale=.6]
     \tikzstyle{Bl}=[draw,circle,blue,scale=.7]
     \tikzstyle{R}=[draw,circle,fill=red,scale=.7]
     \tikzstyle{bl}=[draw,circle,fill=black,scale=.2]
     \tikzstyle{bbc}=[draw,circle,fill=black,scale=.75]
     \tikzstyle{bbcs}=[draw,circle,fill=black,scale=.5]
     \tikzstyle{rc}=[circle,fill=red,scale=.6]
     \tikzstyle{wc}=[draw,circle,scale=.75]
\usepackage{array} 
\usepackage{multirow} 
\usepackage{colortbl} 
\usepackage{cellspace}
\usepackage{xcolor}   

\newcommand{\xdasharrow}[2][->]{
\tikz[baseline=-\the\dimexpr\fontdimen22\textfont2\relax]{
\node[anchor=south,font=\scriptsize, inner ysep=1.5pt,outer xsep=2.2pt](x){#2};
\draw[shorten <=3.4pt,shorten >=3.4pt,dashed,#1](x.south west)--(x.south east);
}
}

\def\^{\wedge}

\def\C{\mathbb{C}}

\def\Z{\mathbb{Z}}

\def\cC{{\mathcal C}}

\newcount\hour \newcount\minute
\hour=\time \divide \hour by 60
\minute=\time
\def\now{%
\ifnum \hour<13
  \ifnum \hour=0 \advance \hour by 12 \number\hour:\else \number\hour:\fi%
     \ifnum \minute<10 0\fi%
     \number\minute%
\ A.M.%
\else \advance \hour by -12 \number\hour:%
  \ifnum \minute<10 0\fi%
  \number\minute%
  \ P.M.%
\fi%
}

\usepackage{tikz}
\usetikzlibrary{arrows}
\usetikzlibrary{shapes.geometric,calc,arrows, positioning,shapes.misc,decorations.markings}
\tikzset{
  big arrow/.style={
    decoration={markings,mark=at position 1 with {\arrow[scale=2,#1]{>}}},
    postaction={decorate},
    shorten >=0.4pt},
  big arrow/.default=black}
\usetikzlibrary{external}
\usetikzlibrary{positioning}
\usetikzlibrary{calc}
\tikzset{gauge-node/.style={shape=circle, draw, minimum width=.6cm}}

\pgfdeclarelayer{edgelayer}
\pgfdeclarelayer{nodelayer}
\pgfsetlayers{edgelayer,nodelayer,main} 
\tikzstyle{none}=[inner sep=0pt] 

\tikzstyle{NodeCross}=[draw, shape=circle, cross out, inner sep=0pt, minimum size=6pt,line width=0.25mm]
\tikzstyle{Circle}=[draw, shape=circle, black, inner sep=0pt, minimum size=6pt]
\tikzstyle{rtriangle}=[fill=black, regular polygon, regular polygon sides=3, rotate=90, inner sep=0pt, minimum size=8pt]
\tikzstyle{ltriangle}=[fill=black, regular polygon, regular polygon sides=3, rotate=270, inner sep=0pt, minimum size=8pt]
\tikzstyle{rtriangleblue}=[fill={rgb,255: red,17; green,160; blue,255}, regular polygon, regular polygon sides=3, rotate=90, inner sep=0pt, minimum size=8pt]
\tikzstyle{ltriangleblue}=[fill={rgb,255: red,17; green,160; blue,255}, regular polygon, regular polygon sides=3, rotate=270, inner sep=0pt, minimum size=8pt]
\tikzstyle{rtrianglegreen}=[fill={rgb,255: red,69; green,255; blue,28}, regular polygon, regular polygon sides=3, rotate=90, inner sep=0pt, minimum size=8pt]
\tikzstyle{ltrianglegreen}=[fill={rgb,255: red,69; green,255; blue,28}, regular polygon, regular polygon sides=3, rotate=270, inner sep=0pt, minimum size=8pt]
\tikzstyle{Uprtriangle}=[fill=black, regular polygon, regular polygon sides=3, rotate=0, inner sep=0pt, minimum size=8pt]
\tikzstyle{Downltriangle}=[fill=black, regular polygon, regular polygon sides=3, rotate=180, inner sep=0pt, minimum size=8pt]
\tikzstyle{rtriangleAmber}=[fill={rgb,255: red, 191; green, 144; blue, 63}, regular polygon, regular polygon sides=3, rotate=90, inner sep=0pt, minimum size=8pt]
\tikzstyle{UprtriangleViolett}=[fill={rgb,255: red,255; green,0; blue,0}, regular polygon, regular polygon sides=3, rotate=0, inner sep=0pt, minimum size=8pt]

\tikzstyle{Downltriangle}=[fill=black, regular polygon, regular polygon sides=3, rotate=180, inner sep=0pt, minimum size=8pt]
\tikzstyle{UpRighttriangle}=[fill=black, regular polygon, regular polygon sides=3, rotate=45, inner sep=0pt, minimum size=8pt]
\tikzstyle{UpLefttriangle}=[fill=black, regular polygon, regular polygon sides=3, rotate=315, inner sep=0pt, minimum size=8pt]
\tikzstyle{DownRighttriangle}=[fill=black, regular polygon, regular polygon sides=3, rotate=135, inner sep=0pt, minimum size=8pt]
\tikzstyle{DownLighttriangle}=[fill=black, regular polygon, regular polygon sides=3, rotate=225, inner sep=0pt, minimum size=8pt]

\tikzstyle{Star}=[draw, shape=star, fill=black, star points=8, inner sep=0pt, minimum size=8pt]

\tikzstyle{DashedLine}=[-, densely dashed, line width=0.25mm]
\tikzstyle{DashedLineBrown}=[-, densely dashed, line width=0.25mm, draw={rgb,255: red,155; green,103; blue,51}]
\tikzstyle{DashedLineFall}=[-, densely dashed, line width=0.25mm, draw={rgb,255: red,195; green,0; blue,0}]
\tikzstyle{DashedLineViolett}=[-, densely dashed, line width=0.25mm, draw={rgb,255: red,139; green,41; blue,148}]
\tikzstyle{DottedLine}=[-, dotted, line width=0.25mm]
\tikzstyle{BlueLine}=[-, fill=none, draw={rgb,255: red,17; green,160; blue,255}, line width=0.25mm]
\tikzstyle{GreenLine}=[-, fill=none, draw={rgb,255: red,69; green,255; blue,28}, line width=0.25mm]
\tikzstyle{RedLine}=[-, draw={rgb,255: red,191; green,0; blue,0}, fill=none, line width=0.25mm]
\tikzstyle{DashedLineRed}=[-, densely dashed, fill=none, draw={rgb,255: red,191; green,0; blue,0}, line width=0.25mm]
\tikzstyle{ThickLine}=[-, line width=0.25mm]
\tikzstyle{ViolettLine}=[-, draw={rgb,255: red,132; green,60; blue,191}, fill=none, line width=0.25mm]
\tikzstyle{ViolettDashedLine}=[-, densely dashed, draw={rgb,255: red,132; green,60; blue,191}, fill=none, line width=0.25mm]
\tikzstyle{AmberLine}=[-, draw={rgb,255: red,191; green,144; blue,63}, fill=none, line width=0.25mm]
\tikzstyle{DashedRedThick}=[-, densely dashed, fill=none, draw={rgb,255: red,191; green,0; blue,0}, line width=0.40mm]
\tikzstyle{DashedBlueThick}=[-, densely dashed, fill=none, black, line width=0.40mm]

\tikzstyle{bulk}=[fill=blue, draw=none, shape=circle,  minimum size=1pt]
\tikzstyle{face}=[fill={rgb,255: red,255; green,128; blue,0}, draw=none, shape=circle,minimum size=1pt]
\tikzstyle{vertex}=[fill=red, draw=none, shape=circle,minimum size=1pt]

\tikzstyle{lattice line}=[-, draw={rgb,255: red,128; green,128; blue,128}, line width = 0.5mm]
\tikzstyle{chamber line}=[-, draw={black}, densely dashed, line width  = 0.7mm]

\newcommand{\qq}{\mathfrak{q}}

\makeatother

\begin{document}

\title{MTC$[M_3, G]$:
3d Topological Order Labeled by Seifert Manifolds}

\author{Federico Bonetti}

\affiliation{Department of Mathematical Sciences, Durham University, Durham, DH1 3LE, United Kingdom}

\author{Sakura Sch\"afer-Nameki}
\author{Jingxiang Wu}

\affiliation{Mathematical Institute, University
of Oxford, Woodstock Road, Oxford, OX2 6GG, United Kingdom}

\begin{abstract} 
\noindent 
We propose a correspondence between topological order in 2+1d and Seifert three-manifolds together with a choice of ADE gauge group $G$. Topological order in 2+1d is known to be characterized in terms of  modular tensor categories (MTCs), and we thus propose a relation between MTCs and Seifert three-manifolds. The correspondence defines for every Seifert manifold and choice of $G$ a fusion category, which we conjecture to be modular whenever the Seifert manifold has trivial first homology group with coefficients in the center of $G$.  
The construction determines the spins of anyons and their S-matrix, and provides a constructive way to determine the R- and F-symbols from simple building blocks. 
 We explore the possibility that this correspondence provides an alternative classification of MTCs, which is put to the test by realizing all  MTCs (unitary or non-unitary) with rank $r\leq 5$ in terms of Seifert manifolds and a choice of Lie group $G$. 

\end{abstract}



\maketitle

\tableofcontents

\section{Introduction}

\noindent
A tantalizing feature of three spacetime dimensions is the existence of non-trivial topological order. Mathematically 3d topological order is characterized by a modular tensor category (MTC) \cite{Moore:1988qv}, for reviews see e.g.~\cite{Kitaev:2005hzj, Wen:2015qgn, EGNO, Simon:2023hdq}. 
In particular, this includes objects known as anyons, which are topological lines, along with data about their spins, braiding, and fusion.
A particularly important piece of information is the modular data, encoded in the S- and T-matrices, which in turn specify the braiding and spins of anyons. 
Unfortunately it was recently shown that this modular data alone is insufficient to uniquely determine an MTC \cite{mignard2021modular}\footnote{Although the first such ambiguity occurs at a relatively high rank, 49.}. To pin down a 3d topological order uniquely, requires specifying the F- and R-symbols as well. 

Nevertheless, a first order classification of topological order includes determining all the modular data. 
A known set of necessary conditions on the S- and T-matrices exist, which ensure that they correspond to the modular data of an MTC. Solutions to these (not necessarily sufficient) conditions were recently determined until rank $r=11$ \cite{Ng:2023wsc}, with earlier results for lower rank in \cite{ostrik2002fusion, ostrik2005premodular, Rowell:2007dge, Wen:2015qgn,bruillardClassificationModularCategories2016,ngReconstructionModularData2022}. But a conceptual and completely explicit classification is still very much wanting. 

In this paper, we shall not find a solution to this formidable problem. However we will propose an alternative approach to studying and potentially classifying 
 MTCs, by relating them to so-called Seifert three-manifold  and a choice of gauge group $G$ (usually chosen to be a Lie group of ADE type). 
 In short, we propose a map 
\be
(\text{Seifert }M_3,  G_{ADE})  \ \rightarrow \ (S, T) \text{ of }\text{MTC}\left[M_3,  G_{ADE}\right] \,,
\ee
i.e. we extract the modular S- and T-matrices, though we will not address the ambiguity in terms of the modular data. However, we will show that this framework surprisingly maps out huge swaths of the set of known MTCs, and we shown that at least to rank  $r= 5$ there is a realization of all known (unitary and non-unitary) MTCs within this framework. 
 
 The proposal is first of all quite ad hoc. Some initial inspiration may have come from the 3d-3d correspondence \cite{Dimofte:2011ju} and subsequently the conjectures in \cite{Cho:2020ljj, Cui:2021lyi}. However ultimately we make no claim of a connection at this point to the 3d-3d correspondence.  
 
 The conjectures of the present work can be formulated simply 
 as a direct characterization of topological order (including non-unitary TQFTs), by a map from the data of the Seifert manifold and a choice of ADE Lie group, to a set of anyons and their T- and S-matrices. The correspondence passes some simple checks, e.g.~that similarity transformations of Seifert manifolds map to the same modular data, and furthermore, we provide a (conjectured) criterion for when a three-manifold gives rise to modular data in terms of the condition $H_1(M_3, \cZ_G) =0$, where $\cZ_G$ is the center of the gauge group $G$. 

 To briefly summarize the correspondence, the main ingredient is the data of a Seifert manifold, which is comprised of $n$ pairs of coprime integers $(p_i,q_i)$, each of which characterizes a singular fiber.
 We consider flat $G_\C$ connections on such a three-manifold, satisfying certain conditions, called {\bf anyonic flat connections}. These will play the role of the anyons in the dual MTC. For each fiber parametrized by $(p_i,q_i)$ we associate a pre-modular category $\mathcal{C}(\mathfrak{sl}(N),p_i, q_i)$. 
 The fusion category associated to $M_3$ is obtained by taking a (graded) Deligne product of $\mathcal{C}(\mathfrak{sl}(N),p_i, q_i)$. We provide strong evidence that that this is modular if and only if $H_1(M_3, \cZ_G) =0$. Restricting to this set of three-manifolds we map out the low rank MTCs and find a complete list up to rank $r\leq 5$.

The proposal of this work may superficially have similarities  with the one in \cite{Cho:2020ljj,Cui:2021lyi}, which was  further developed in \cite{Cui:2021yes}, where also a curious connection between three-manifolds and 3d topological order was conjectured. In short the proposal {there} is that compactifying the 6d $(2,0)$ supersymmetric, super-conformal field theory  with gauge algebra $\mathfrak{su}(2)$ on a three-fibered Seifert manifold $M_3$, gives rise to a UV $\mathcal{N}=2$ supersymmetric 3d gauge theory $T[M_3]$, which conjecturally flows in the IR to a  gapped phase.  
This approach yields all MTCs up to rank $r=4$, but fails to produce all MTCs at higher rank (it fails to reproduce $r=5$ models, concretely the models 1., 2., 21.-26. in table \ref{tab:Rank5} do not have a realization in the proposal \cite{Cho:2020ljj,Cui:2021lyi}). The connection between three-manifolds $M_3$ and MTCs is made -- in the spirit of the 3d-3d-correspondence \cite{Dimofte:2011ju} -- via the Chern-Simons invariant and Reidemeister Torsion of  flat $SL(2,\C)$  connections on $M_3$.  E.g.~in determining all rank 4 models, it is crucial for \cite{Cho:2020ljj} to include models with non-trivial $H_1(M_3, \cZ_G)$ (which are thus not automatically modular), and require a gauging of a 1-form symmetry given by $H_1(M_3, \cZ_G)$, and completion to a modular tensor category. 
Although the current work was motivated by this connection, a naive extension seems unclear (in particular the question of computing the torsion for higher number of fibers or higher rank groups is at best ill-defined). In addition, for a generic $M_3$, the 3d $\mathcal{N}=2$ SQFT is not gapped in the infrared. It is clear that, without a deeper understanding of the 3d-3d correspondence and the RG flow from the UV 3d $\mathcal{N}=2$ SQFT to the IR phase, a  precise map between $T[M_3]$ and MTCs is currently out of reach. 

The connection between Seifert manifolds and MTCs that we propose here is a priori  {\it{\textbf{distinct}}} from the above, and in particular does not hinge on the 3d-3d correspondence. We will at this point simply provide a dictionary between the data Seifert manifolds and  Lie groups $G$, and MTCs (not necessarily unitary). Crucially we relate triviality of $H_1(M_3, \cZ_G)$ with modularity, and only consider Seifert manifolds with $H_1(M_3, \cZ_G)=0$. 
In particular this means that even for those MTCs that have a realization within the framework of \cite{Cho:2020ljj}, many will have non-trivial $H_1$, and thus our proposed realization in terms of Seifert manifolds is quite different, as exemplified  in  these concrete examples.

Finally, one might ask how our proposal differs from the realization of MTCs in terms of Chern-Simons (CS) theories (and related cosets) \cite{Moore:1988qv,Wen:2015qgn}.  
The construction of MTC$[M_3, G]$ is based on building blocks, which are pre-modular categories. These are associated to building blocks (fibers) of the Seifert manifold $M_3$. The MTC is obtained as a graded Deligne product of these building blocks. Although some of the building blocks may be of CS-type, the graded Deligne product is not.  

Furthermore, the construction we propose also provides a systematic way to compute the R- and F-symbols (using the graded Deligne product on the pre-modular categories). As the building blocks have well-known R- and F-symbols, we view this as an advantage of this construction, as it allows bootstrapping this data for MTCs which are otherwise notoriously hard to compute \cite{Rowell:2007dge}.

\noindent
{\bf Summary.} Let us summarize the construction: 
A Seifert manifold $M_3$ is characterized in terms of $n$ fibers (see below). We associate to each fiber a pre-modular tensor category. 
This is graded by the center of $G$: $\cZ_G$. If $H_1 (M_3, \cZ_G) =0$, the $\cZ_G$-graded Deligne product of these pre-modular categories is modular and defines our MTC$[M_3, G]$: 
\begin{itemize}
\setlength\itemsep{-0.1em}
\item Modularity of this is equivalent to vanishing of $H_1(M_3, \cZ_G)$ where $\cZ_G$ is the center of the gauge group $G$. 
\item The choice of building blocks and the gluing is determined by the data of the Seifert three-manifold $M_3$. 
\item Equivalence between different presentations of the three-manifold give equivalent MTCs.
\item Beyond providing the T- and S-matrices from the Seifert data and information of $G$, the construction also provides a concrete way to compute R- and F-symbols. 
\item This results in all MTCs (not necessarily unitary!) for rank $r\leq 5$ for which we list $(M_3, G)$ in appendix 
\ref{app:MTCTables}, and provide a model for each of the MTCs in  appendix E of \cite{Ng:2023wsc}. 
\end{itemize}

The structure of the paper is as follows: we provide background on Seifert manifolds and flat connections in section \ref{sec:Seifert}, and we introduce the notion of an anyonic flat connection, that will be central to our correspondence.
A summary of MTCs can be found in section \ref{sec:MTCs}, where we also introduce a graded Deligne product, which will be the key tool for the contruction based on Seifert data. 
The main proposal relating these two things is explained in section \ref{sec:Meat}. We put the conjecture that this may provide a comprehensive list of MTCs to test in section \ref{sec:AllMTC}, with tables of all MTCs up to rank 5 in the appendix \ref{app:MTCTables}. The precise map between anyonic flat connections and anyons is detailed in appendix \ref{app_CS_match}.

\section{Seifert Manifolds and Flat Connections}
\label{sec:Seifert}

\subsection{Seifert 3-manifolds}

A Seifert manifold $M_3$ is a 
closed, connected, smooth 3-manifold that admits a smooth circle action, see \cite{OrlikBook,Neumann1978}.
The base of the fibration is 
a two-dimensional orbifold, given by a genus $g$ Riemann surface $\Sigma_{g,n}$
with $n$ distinct marked points.
In this work we are only interested in
orientable Seifert fibered spaces with orientable base.
A Seifert fibration is specified by the data
$b$, $g$ $\{(p_k,q_k)\}_{k=1}^{n}$,
where $g$, $b$, $p_k$, $q_k$ are integers,
$g\ge 0$, $p_k \ge 2$,
$\gcd(p_k, q_k) = 1$. 
The order of the pairs
$(p_k,q_k)$ is immaterial and there may be repeated
pairs. We will restrict in this paper to $g=b=0$ and denote the associated 3-manifold by
\be 
M_3=  [\{(p_k,q_k) \}_{k=1}^n] \,.
\ee 
The Seifert data in this paper
is not  normalized. 
The fundamental group
of the Seifert manifold
$M_3=[\{(p_k,q_k) \}_{k=1}^n]$
has generators $\{x_k\}_{k=1}^n$ and $h$ subject to 
 relations 
\be 
x_k^{p_k} h^{q_k}=1 \ , \quad 
x_1 x_2 \dots x_n=1 \ , \quad 
\text{$h$ is central} \  .
\ee 
Hurewicz theorem
implies that 
\be 
H_1(M_3;\mathbb Z) \cong {\rm coker}\, M \ , 
\ee 
where $M$ is the following $(n+1)\times (n+1)$ matrix,
\small
\be 
M = \begin{pmatrix}
1 & 1 & \cdots & 1 & 0 \\
p_1 & 0 & \cdots & 0 & q_1 \\
0 & p_2  & \cdots & 0 & q_2 \\ 
\vdots  &&&& \vdots \\
0 & 0  & \cdots & p_n & q_n \\ 
\end{pmatrix}  \ . 
\ee 
\normalsize
When $n=3$ this simplifies to
$H_1(M;\mathbb Z) \cong \mathbb Z_K \oplus
\mathbb Z_L$, where
\be 
K = \gcd(p_1, p_2,   p_3) \ , \ \ 
L = \frac{p_1 p_2  p_3}{ K} \, \bigg|
\sum_{k=1}^3\frac{q_k}{p_k}  
\bigg| \ .
\ee 
The main focus of this paper will be on $M_3$ with trivial $H_1(M_3, \cZ_G)$, where $\cZ_G$ will be the center of the gauge group of ADE type.

\subsection{Flat Connections}

Fix the Seifert manifold
$M_3=[\{(p_k,q_k) \}_{k=1}^n ]
$
and the complex Lie group
$SL(N,\mathbb C)$. 
A flat $SL(N,\mathbb C)$ connection on $M_3$
is the same as a group homomorphism
\be 
\rho: \pi_1(M_3) \rightarrow SL(N,\mathbb C) \, , 
\ee 
up to conjugation.
A flat $SL(N,\mathbb C)$ connection on $M_3$ is said to be \textbf{irreducible}
if it has a finite stabilizer group 
\be 
{\rm Stab}(\rho) = \{ g\in SL(N,\mathbb C)  | 
g \rho(x) = \rho(x) g ,  \forall x\in \pi_1(M_3) 
\} \,.
\ee 
Assume $\rho$ is an irreducible
flat $SL(N,\mathbb C)$ connection on $M_3$.
Then, the following properties hold.
Firstly, the matrix
$\rho(h)$ must be an element
in the center of $SL(N,\mathbb C)$,
\be  \label{eq_rho}
\rho(h) = e^{2\pi i \ell/N} \mathbb I_N \ , \quad 
\ell \in \{0,\dots,N-1\} \ .
\ee 
Secondly, the matrix $\rho(x_k)$ is diagonalizable
for each $k=1,\dots,n$.
Finally, at most $n-3$
out of the $n$ matrices
$\rho(x_k)$  can be
a multiple of the identity matrix. 
These properties of
irreducible flat connections
are established in \cite{FuturePaper},
where we count them systematically
for various values of $N$, $n$.

\subsection{Anyonic Flat Connections}

We consider a particular type of irreducible
flat $SL(N,\mathbb C)$ connection, which we call 
\textbf{anyonic}, if for each $k=1,\dots, n$, the 
eigenvalues of the matrix $\rho(x_k)$ are all distinct. 
It is this set of flat connections that will 
be relevant in formulating a correspondence to MTCs. 

The moduli space of anyonic $SL(N,\mathbb C)$
connections on $M_3=[\{(p_k,q_k) \}_{k=1}^n]$ generically
consists of multiple connected components.
The matrix $\rho(h)$ and the eigenvalues
of the matrices $\rho(x_k)$ are the same
for all connections $\rho$
in the same connected component. 
We have already parametrized
$\rho(h)$ in \eqref{eq_rho}.
Next, we seek a convenient
parametrization of the eigenvalues of
$\rho(x_k)$ compatible
with the relation
$\rho(x_k)^{p_k} \rho(h)^{q_k} 
= \mathbb I_N$ and with the requirement
that all eigenvalues of $\rho(x_k)$ be
distinct.
To this end,
we write 
\be  \label{eq_rho_x_diag}
\rho(x_k) \sim \diag(e^{2\pi i a_k^{(1)}} , \dots , e^{2\pi i a_k^{(N)}}) \ , 
\ee 
where $\sim$ is equality up to
conjugation in $SL(N,\mathbb C)$ and
the rational numbers
$a_k^{(I)}$, $I=1,\dots,N$, satisfy
\be 
\!\!\! a_k^{(1)} < a_k^{(2)} < \dots < a_k^{(N)} < a_k^{(1)}+1 \ , 
\ \ 
\sum_{I=1}^N a_k^{(I)}= 0 \ .
\ee 
This is up to conjugation the most general diagonal matrix with distinct entries that are roots of unity. 
The relation $\rho(x_k)^{p_k} \rho(h)^{q_k} 
= \mathbb I_N$ implies
that the $a_k^{(I)}$ take the form
\be  \label{eq_a_param_with_m}
a_k^{(I)} = \frac{1}{p_k} \bigg( \nu_k^{(I)} - \frac{\ell q_k}{N} \bigg) \ , \quad   I=1,\dots, N \ ,
\ee 
where $\nu_k^{(I)}\in \mathbb Z$ with 
$\sum_{I=1}^N \nu_k^{(I)} = \ell q_k$. 
Here the
integers $\nu$ are unique
once we demand 
$\nu^{(1)}_k < \dots < \nu_k^{(N)} < \nu^{(1)}_k+p_k$
together with
$\sum_{I=1}^N \nu_k^{(I)} = \ell q_k$.

Our discussion above can be summarized
by stating that,
to each connected component
in the space of anyonic
$SL(N,\mathbb C)$
flat connections on $M_3$,
we can associate an
element in the finite set
\be \label{eq_R_set}
{\mathcal R}^N_{\{ (p_k,q_k)\}_{k=1}^n}
\!\!\!
:=
\!\!
\bigsqcup_{\ell=0}^{N-1}
\left \{ \!
\begin{array}{l}
\nu^{(I)}_k \in \mathbb Z \ , \
1\le I \le N \ , \
1 \le k \le n
\\[1.5mm]
\nu^{(1)}_k < \dots < \nu_k^{(N)} < \nu^{(1)}_k+p_k \\[1.5mm]
\sum_{I=1}^{N} \nu^{(I)}_k = \ell q_k
\end{array}
\! \!
\right \}
\ .
\ee 
On the other hand, it is a highly nontrivial question which element in ${\mathcal R}^N_{\{ (p_k,q_k)\}_{k=1}^n}
\!\!\!$ defines an anyonic flat connection.

For given $N$ and Seifert data
it is possible to set up an explicit
numerical counting problem of
anyonic $SL(N,\mathbb C)$ flat connections.
This will be reported in \cite{FuturePaper}.
The findings of such numerical investigations
indicate that every point in the set
\eqref{eq_R_set} is associated to
a connected component
in the space of anyonic
flat connections. We then propose the following:

\begin{conj}
The connected components of the anyonic $SL(N, \C)$ flat connections 
on the Seifert
manifold
$M_3= [  \{(p_k,q_k) \}_{k=1}^n]$
are in 1-1 correspondence with points in the finite set
${\mathcal R}^N_{ \{ (p_k,q_k)\}_{k=1}^n}$
in \eqref{eq_R_set}.
\end{conj}

If $\rho$ is a flat $SU(N)$ connection on $M_3$, the value of the classical Chern-Simons (CS) invariant $\mathrm{CS}(\rho)$
of $\rho$ is~\cite{nishi}
\begin{equation} \label{eq:CSformula}
{\rm CS}(\rho)  = \frac 12 \sum_{k=1}^n
\bigg[
p_k r_k {\rm Tr} X_k^2
- q_k s_k {\rm Tr}H^2
\bigg] \!\!\!\!\! \mod \mathbb Z \, ,
\end{equation}
where the integers $r_k$, $s_k$ satisfy $p_k s_k - q_k r_k  = 1$ and the traceless matrices $X_k$,
$H$ are  given by
\be  \label{eq_def_X_and_H}
\ba 
X_k &= {\rm diag}(a_k^{(1)} , \dots , a_k^{(N)}) \ , \\ 
H &={\rm diag}\left(
\tfrac{\ell}{N},\dots,\tfrac{\ell}{N}, - \tfrac{(N-1)\ell}{N}
\right) \ . 
\ea 
\ee 
Since $\mathrm{CS}(\rho)$
depends on the eigenvalues of
$\rho(h)$, $\rho(x_k)$,
it is constant in each connected
component of the space of flat connections.
Crucially, the group we consider is the complexification
$SL(N,\mathbb C)$ and there is an important question that we will not address, whether the CS invariants on Seifert
manifolds differ when the group is complexified. For $SU(2)$ and $SL(2,\C)$ we know that the invariants agree. It would be interesting to analyze this question in general.

\section{3d Topological Order and MTCs}
\label{sec:MTCs}

\noindent{\bf MTC and RFC.}
The second important ingredient in our setup are the so-called {\bf modular tensor categories} (MTCs), characterizing 3d bosonic topological field theories, or bosonic topological order. Mathematically, an MTC $\cC$ is a fusion category, whose objects are topological lines, or anyons. MTCs  have finitely many simple objects $X_j$ with $j\in \mathcal{I}$ for an index set $\mathcal{I}$, so that any object is a direct sum of $X_i$. The rank is defined as the cardinality $|\mathcal{I}|$ of the set $\mathcal{I}$. The fusion coefficients $N_{ij}^k \in \mathbb{N}$ appear in the tensor product
\be
X_i \otimes X_j  = \bigoplus_k N_{ij}^k X_k 
\ee
and satisfy various consistency conditions. 
We will label an MTC $\mathcal{C}$ (tt is known that the modular data $(S,T)$ does not uniquely determine an MTC \cite{mignardModularCategoriesAre2021}) in terms of the following data:
\begin{itemize}
\item Rank $r=|\mathcal{I}| \equiv |\mathcal{C}|$, which is the number of simple topological lines, i.e.~simple anyons. 
\item The spin $h_X$, or equivalently the twist $\theta_X = \exp(2\pi ih_X)$  of the line $X \in \cC$, which is a root of unity.  The twists form a diagonal matrix $T_{ij} = \delta_{ij} \theta_i$, $i,j = 1,\dots, r$. We follow the convention that the spin of the identity line is $0$, or equivalently $T_{11}=1$.
\item The S-matrix $S_{ij}$ is a unitary \emph{non-degenerate} symmetric matrix that encodes the braidings of two anyons $X_i$ and $X_j$. 
\end{itemize}

If we drop the requirement of having a non-degenerate S-matrix, one instead has a ribbon fusion category (RFC), also known as a pre-modular category. For an MTC to be \emph{unitary}, one simple necessary condition is to require all quantum dimensions $d_i\equiv S_{i0}/S_{00}$ to be real numbers greater or equal to $1$.\\

\noindent{\bf The MTC $\mathcal{C}(\mathfrak{sl}(N), p, \mathfrak{q})$.} One of the most important approaches to MTCs is through the representations of quantum groups at roots of unity. In particular, the semisimple quotient of the category of finite dimensional modules of quantum group $U_{\qq}(\mathfrak{sl}(N))$ at roots of unity gives rise to an (not necessarily modular) RFC, denoted as $\mathcal{C}(\mathfrak{sl}(N), p,\qq)$ where $\qq^2$ is a root of unity of order $p$ \cite{chariGuideQuantumGroups1995,sawinQuantumGroupsRoots2003, Rowell:2005hv, schopierayLieTheoryFusion2018}.

\begin{figure}[htbp]
    \centering
    \includegraphics[width = 0.7 \columnwidth]{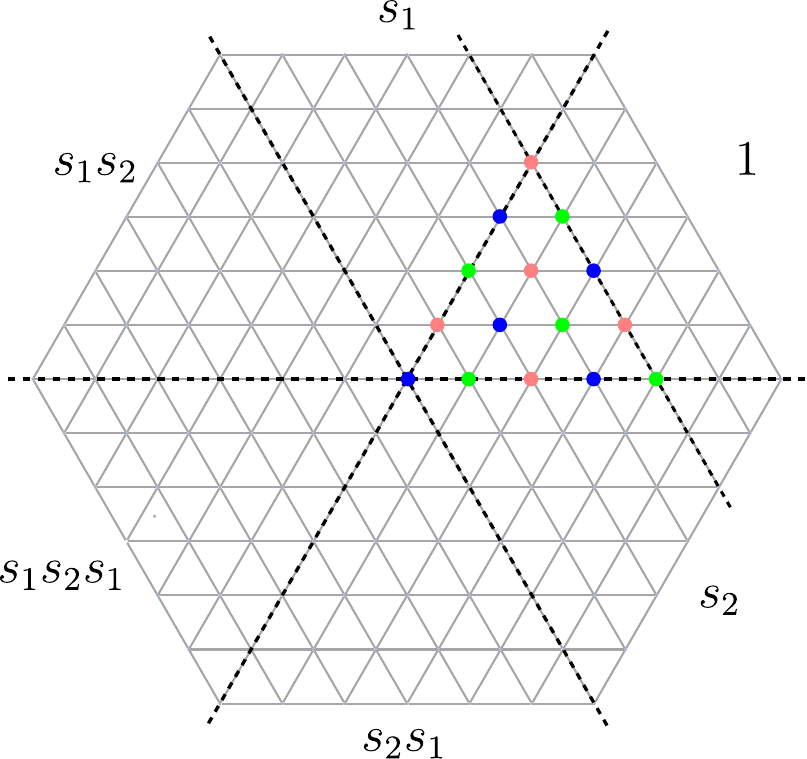}
    \caption{Fundamental affine Weyl alcove  ${\Delta}_{3,7}$ for $\mathfrak{sl}(3)$ with $\phi(\lambda) = 0,1,2$ for blue, green and orange respectively. Here $s_i$ are Weyl reflections, which label all the different alcoves. We will restrict to the fundamental one (labeled by 1).}
    \label{fig:weightSU3}
\end{figure}

We recall that the fundamental affine Weyl alcove  ${\Delta}_{N,p}$
of the affine Lie algebra
$\mathfrak{sl}(N)$
at level $p-N$
consists of those weights
with integral Dynkin
labels $[\lambda^{(1)} , \dots ,
\lambda^{(N-1)}]$
in the set 
\be \label{eq_def_Delta}
    {\Delta}_{N,p}:= 
    \!\Big\{ 0\leq \lambda ^{(I)} \leq p-h^\vee, \, I = 0,\dots, N-1 \Big\}  \ , 
\ee 
where $ \lambda^{(0)} := (p-h^\vee) - \sum_{I=1}^{N-1} \lambda^{(I)}$, $h^\vee =N$.
Each $\lambda \in  \Delta_{N,p}$ has an $N$-ality
defined by the map
\be \label{eq_def_phi}
 \phi(\lambda) :=
\sum_{I=1}^{N-1} I \lambda^{(I)}
\mod N \ . 
\ee 
We depict ${\Delta}_{N,p}$
for $N=3$, $p=7$ in figure
\ref{fig:weightSU3}, where the colors represent the values of $\phi(\lambda)$. Fundamental alcoves for general simply-laced $\mathfrak{g}$ will be discussed in appendix \ref{app:Alcove}.

After these preliminaries,
let us summarize some salient features of the RFC $\mathcal{C}(\mathfrak{sl}(N), p,\qq)$: 
\begin{itemize}
    \item Simple objects are labelled by the weights in the fundamental affine Weyl alcove ${\Delta}_{N,p}$ of the affine Lie algebra ${\mathfrak{sl}(N)}$ at level $p-N$,
    defined in \eqref{eq_def_Delta}.
    One can see immediately that the rank $|\mathcal{C}(\mathfrak{sl}(N), p, \qq)|$ is independent of $\qq$ and can be found from the coefficient of $x^{p-N}$ in the power series expansion of $(1-x)^{-N}$. 
    \item The fusion rules are given by the familiar fusion coefficients of affine representations \cite{DiFrancesco:1997nk}.
    \item For any two dominant weights $\lambda, \mu \in {\Delta}_{N,p}$, the S-matrix element is \cite{schopierayLieTheoryFusion2018}
\be\label{SmatSchopi}
S_{\lambda {\mu}}=\frac{\sum_{\sigma \in W}\epsilon(\sigma) \qq^{2\langle \sigma({\lambda}+\rho),\mu+\rho\rangle}}{\sum_{\sigma \in W}\epsilon(\sigma) \qq^{2\langle \sigma(\rho),\rho\rangle}} \,,
\ee
where $\rho$ is the half sum of the positive roots and $\epsilon(\sigma)$ denotes the sign of the Weyl group element $\sigma\in W$ defined by the Bruhat order.
    \item The twists are
    \begin{equation} \label{eq_twist}
    \theta_{\lambda} = \qq^{\langle \lambda, \lambda+2\rho\rangle }\ . 
    \end{equation} 
\end{itemize}
In addition, to fully characterize an MTC, one needs to provide $F$- and $R$-symbols subject to pentagon and hexagon identities. For $\mathfrak{g} = \mathfrak{sl}(2)$, this can be found in~\cite{kauffmanTemperleyLiebRecouplingTheory1994}.

We observe that the S- and T-matrices
contain fractional powers of $\qq$.
This is because the inner product
$\langle \lambda , \mu \rangle$ is quantized in
units of $1/N$ if $\lambda$, $\mu$ are integral. In order to write unambiguous
expressions for $S$ and $T$, we parametrize $\qq$~as
\be 
\qq = \mathfrak{s}^N \ . 
\ee 
Then, $S_{\lambda \mu}$ is a ratio of polynomials in the
variable $\mathfrak s^2$, while $\theta_\lambda
= \mathfrak s^{M(\lambda)}$
with $M(\lambda)$ a nonnegative integer.
If $N$ is even, $M(\lambda)$ can be even or odd,
depending on $\lambda$.
If $N$ is odd, $M(\lambda)$ is even for any $\lambda$.
It follows that, for $N$ odd,
all entries of the S- and T-matrix
are functions of~$\mathfrak s^2$.
In all cases, we require
that $\mathfrak s^{2N}$ has order
$p$ to ensure that
$\qq^2$ has order
$p$.

We emphasize that the existence of S- and T-matrices does not necessarily imply that the category is modular, i.e.~for a generic root of unity $\qq$, $\mathcal{C}(\mathfrak{sl}(N),p, \qq)$ is not necessarily modular. 
Bruguieres \cite{bruguieres2000categories} 
shows that $\mathcal{C}(\mathfrak{sl}(N), p, \qq)$ for $\qq=e^{z \pi i / p}$ is modular  if and only if $\operatorname{gcd}(z, N p)=1$.
In particular, when 
\be
\ba
&\qq (k, N) = \exp\left(\frac{\pi i}{k+N} \right): \cr 
&\mathcal{C}(\mathfrak{sl}(N), k+N, \qq(k,N))= \text{Rep}\left(\widehat{\mathfrak{sl}}(N)_k\right) 
\ea
\ee
is known to be an MTC given by the representation category 
$\text{Rep}(\widehat{\mathfrak{sl}}(N)_k)$ 
of the affine Lie algebra, or equivalently the category of Wilson lines in the $\mathfrak{sl}(N)_k$ Chern Simons theory. 
\\

\noindent{\bf Graded Deligne Product.}
The final building block is to define a graded Deligne product $\boxtimes_G$ for RFCs as in \cite{Cui:2021lyi}. Suppose $A$ is an abelian group and $\mathcal{C}=\oplus_{g \in A} \mathcal{C}_g$ and $\mathcal{D}=\oplus_{g \in A} \mathcal{D}_g$ are two $A$-graded RFC's. The $A$-graded product $\boxtimes_{A}$ of $\mathcal{C}$ and $\mathcal{D}$ is defined to be $\mathcal{C} \boxtimes_{A} \mathcal{D}=\oplus_{g \in A}\mathcal{C}_g \boxtimes \mathcal{D}_g$, which is a subcategory of the usual Deligne product $\mathcal{C}\boxtimes \mathcal{D}$.

$\mathcal{C} \boxtimes_{A} \mathcal{D}$  is again a $A$-graded ribbon fusion category, where all of the category data are defined by a straightforward element-wise multiplication. If we write simple objects in a $A$-graded RFC $\mathcal{C}$ as $c_{g,i}\in \mathcal{C}_g$, $i = 1, \dots, |\mathcal{C}_g|$, $\mathcal{C} \boxtimes_{A} \mathcal{D}$ has the data
\begin{itemize}
    \item Simple objects: $c_{g,i}\boxtimes d_{g,j}$, for all $c_{g,i}\in \mathcal{C}_g$, $d_{g,j} \in \mathcal{D}_g$ and $g\in A$.
    \item Rank: $r= \sum_g |\mathcal{C}_g||\mathcal{D}_g|$.
    \item  S-matrix: 
    \begin{align}
        S_{c_{g,i}\boxtimes d_{g,j},c_{h,m}\boxtimes d_{h,n}} \equiv S_{c_{g,i}, c_{h,m}} S_{d_{g,j},d_{h,n}} \,.
    \end{align}
    \item Twists:
    \begin{equation}
    \label{eq_combine_twists}
        \theta_{c_{g,i}\boxtimes d_{h,j}} \equiv \theta_{c_{g,i}}\theta_{d_{h,j}}  \,.
    \end{equation}
\end{itemize}
$F$- and $R$-symbols are defined similarly by a element-wise multiplication of the ones of $\mathcal{C}$ and $\mathcal{D}$ respectively.

More generally, the $A$-graded Deligne product of multiple $A$-graded RFCs can be defined in a similar manner
\begin{equation}
    \mathcal{C}^1 \boxtimes_A\mathcal{C}^2 \boxtimes_A \dots \boxtimes_A \mathcal{C}^n \equiv \oplus_{g\in A} \mathcal{C}^1_g \boxtimes\mathcal{C}^2_g \boxtimes \dots \boxtimes \mathcal{C}^n_g \,,
\end{equation}
where the ordering does not matter since the $A$-graded product is clearly commutative and associative.

However, it has been observed in \cite{Cui:2021lyi} that $A$-graded Deligne product of $A$-graded MTCs may not be an MTC, namely its S-matrix may not be necessarily non-degenerate. It is therefore an interesting question when $A$-graded Deligne products of MTCs  define MTCs. On the other hand, it is possible that the graded Deligne product of non-modular ribbon fusion categories (RFCs) turns out to be modular.

\section{ MTCs from Seifert Manifolds}
\label{sec:Meat}

We are now in a position to discuss
the proposed map from Seifert manifolds
to RFCs. We will present our two main conjectures, discuss the relation with anyonic
flat connections, and present some examples.\\

\noindent{\bf Defining $\mathsf{RFC}(\mathfrak{sl}(N), M_3)$ and $\mathsf{MTC}(\mathfrak{sl}(N), M_3)$.}
Let us fix the
Seifert manifold 
\be 
M_3= [\{(p_k,q_k) \}_{k=1}^n ] \ ,
\ee
and the complex Lie algebra 
$\mathfrak g= \mathfrak{sl}(N)$.
Assume 
\be 
p_k \ge N \ , \quad 
\gcd(N,q_k)=1 \ , \quad k=1,\dots n \ . 
\ee 
To the pair $(\mathfrak{sl}(N),M_3)$
we associate the RFC denoted 
$\mathsf{RFC} ( \mathfrak{sl}(N) , M_3 )$
and  defined
as the $\mathbb Z_N$-graded product of $n$ factors,
\be  \label{eq_RFC_of_M3}
\mathsf{RFC} ( \mathfrak{sl}(N) , M_3 )
:= \mathcal C^1 \boxtimes_{\mathbb Z_N}
\dots \boxtimes_{\mathbb Z_N} \cC^n \ , 
\ee 
where the factor $\cC^k$ associated to the $k$th
exceptional fiber of $M_3$ is the 
$\mathbb Z_N$-graded
RFC
\be  \label{eq_Ck_building_block}
\cC^k = \mathcal C(\mathfrak{sl}(N) , p_k , \qq_k ) \ , 
\ee 
with $\qq_k$  specified as 
\be \label{eq_qqk}
\qq_k = \mathfrak s_k^N \ , \quad 
\left\{ 
\begin{array}{ll}
\mathfrak s_k = e^{2\pi i \frac{x_k}{2p_kN}} & \text{for $N$ even,}
\\
\mathfrak s^2_k = e^{2\pi i \frac{y_k}{p_kN}} &
\text{for $N$ odd,}
\end{array}
\right.
\ee 
where $x_k$, $y_k$ are integers satisfying 
\be  \label{eq_xy_def}
x_k  = q_k^{-1} \!\!\!\!\mod 2p_kN \ , \quad 
y_k = q_k^{-1} \!\!\!\!\mod p_kN \ . 
\ee 
In performing the graded Deligne product (\ref{eq_RFC_of_M3}) we equip $\cC^k$ with the $\Z_N$ grading defined by 
\be \label{eq_grading}
g_k(\lambda) = q_k^{-1}  \phi(\lambda) \; \;\text{mod} \; N  \,.
\ee

Thus, simple objects in
$\mathsf{RFC} ( \mathfrak{sl}(N) , M_3 )$
are labeled by tuples $\boldsymbol{\lambda}=(\lambda_1, \dots, \lambda_n)$ in the set 
\be  \label{eq_def_cS}
{\mathcal S}^N_{\{ (p_k,q_k)\}_{k=1}^n}
\!\!
:=
\bigsqcup_{g=0}^{N-1}
\left\{  \!\!
\begin{array}{l}
\boldsymbol{\lambda} \in  \Delta_{N,p_1} \times 
 \dots  
\times 
 \Delta_{N,p_n} \text{ s.t.} 
\\[1mm] 
 g_k(\lambda_k)= g \!\!\!\mod N 
\end{array}
\!\!
\right\}  \ ,
\ee 
with $g_k(\lambda)$ as defined in \eqref{eq_grading}.

We observe that $\mathsf{RFC} ( \mathfrak{sl}(N) , M_3  )$
may or may not be modular,
depending on $N$ and the Seifert data.
When it is modular, we also use the notation
$\mathsf{MTC} ( \mathfrak{sl}(N) , M_3 )$.

Since the building blocks $\cC^k$ are not necessarily unitary, the resulting MTCs are also not necessarily unitary. We will see that at least at low rank this proposal yields all unitary and non-unitary modular data. 
\\

\noindent{\bf Main conjectures.}
We formulate two main conjectures
on   $\mathsf{RFC} ( \mathfrak{sl}(N) , M_3  )$.
The first conjecture
asserts that, if we start from
distinct 
Seifert data that yield diffeomorphic
3-manifolds, the associated RFCs
have the same modular data.
More precisely:

\begin{conj}
Fix $N\ge 2$.
Let $\{ (p_k,q_k)_{k=1}^n\}$,
$\{ (p_k,q'_k)_{k=1}^n\}$ be Seifert data
with $p_k\ge N$, 
$\gcd(N,q_k)=1$,
$\gcd(N,q_k')=1$, $k=1,\dots, n$,
such that there exist integers
$\{m_k\}_{k=1}^n$ with
\be 
q'_k = q_k + p_k m_k  \ ,\ \ k=1, \dots, n,  \quad 
 \sum_{k=1}^n m_k=0 \ .\label{eq:seiferttransf} 
\ee 
Write 
\be 
M_3 = [\{ (p_k, q_k) \}_{k=1}^n]
\ , \,\,\, 
M_3' =  [\{ (p_k, q'_k) \}_{k=1}^n] \ . 
\ee 
Then 
$\mathsf{RFC} ( \mathfrak{sl}(N) , M_3 )$
and $\mathsf{RFC} ( \mathfrak{sl}(N) , M_3')$
have the same modular data.
\label{conj:equivundergauge}
\end{conj}

We have tested this conjecture in several
examples, by verifying that  
$\mathsf{RFC} ( \mathfrak{sl}(N) , M_3  )$
and $\mathsf{RFC} ( \mathfrak{sl}(N) , M_3' )$
have the same T- and S-matrices
up to a permutation of simple object labels.

Note that a generic Seifert `gauge' transformation \eqref{eq:seiferttransf} does not preserve the assumption $\gcd(N,q_i)=1$, while keeping $b = 0$. For more details see appendix \ref{sec:moreSeifert}. The assumption $\gcd(N,q_i)=1$ and $\gcd(N,q_i')=1$ will be relaxed later. 

Our second main conjecture relates
the topological properties of
$M_3$ to the non-degeneracy of the S-matrix
of   $\mathsf{RFC} ( \mathfrak{sl}(N) , M_3  )$.
More precisely:

\begin{conj}
Fix $N\ge 2$.
Let $\{ (p_k,q_k)_{k=1}^n\}$
be Seifert data
with $p_k\ge N$, 
$\gcd(N,q_k)=1$.
Assume that at most
one of the $p_k$ is equal to $N$.
Set $M_3  = [\{ (p_k, q_k) \}_{k=1}^n]$.
Then
$\mathsf{RFC} ( \mathfrak{sl}(N) , M_3  )$
is modular if and only if
$H_1(M_3;\mathbb Z_N) =0$.
\end{conj}

 In fact for $n=3$ (three-fibers) and $N=2$ ($SL(2,\C)$) this is proven in \cite{Cui:2021lyi}. We have tested this conjecture
is numerous examples including all the rank $r=2, 3, 4, 5$ that we tabulate in the appendices, going beyond $n=3$ and $N=2$.

If we relax the hypothesis 
that at most one of the $p_k$
is equal to $N$, the conjecture
is no longer valid.
For example, 
if $N=2$, $M_3 = [   (2,1),(2,1),(5,1) ]$ then 
$H_1(M_3;\mathbb Z_2) \cong \mathbb Z_2$
but
$\mathsf{RFC} ( \mathfrak{sl}(2) , M_3 )$
is modular. On the other hand, for $H_1$ non-trivial and only one $p_k=N$, we have confirmed in numerous examples that the models are not modular. \\

\noindent{\bf Defining 
$\mathsf{RFC}(\mathfrak{sl}(N) , M_3)$ for $N$ prime relaxing
$\gcd(N,q_k)=1$.}
When $N$ is prime,
we can extend the definition
of $\mathsf{RFC}(\mathfrak{sl}(N) , M_3)$
by allowing Seifert data in which we do not necessarily
have $\gcd(N,q_k)=1$ for every $k=1,\dots,n$.
The form of $\mathsf{RFC}(\mathfrak{sl}(N) , M_3)$
is the same as in
\eqref{eq_RFC_of_M3},
with building blocks as in 
\eqref{eq_Ck_building_block}.
For each building block
\eqref{eq_Ck_building_block} we have to
specify: (i) the value of
$\qq_k$; (ii) 
the grading function $g_k(\lambda)$,
which might differ from \eqref{eq_grading}.
We now describe these assignments in the cases
$N=2$ and $N$ an odd prime.
The case $N=2$ has been first
discussed in~\cite{Cui:2021lyi}.

If $N=2$, 
the value of $\qq_k$ is 
$\qq_k = \mathfrak s_k^2$
with
\be \label{eq_qqk_N2}
\mathfrak s_k = 
\left\{
\begin{array}{ll}
e^{2\pi i \frac{x_k}{4 p_k }}
&
\text{if $q_k$ odd} \ , \\
e^{2\pi i \frac{\xi_k}{4 p_k }}
&
\text{if $q_k$ even} \ , 
\end{array}
\right.   
\ee
where 
the integers $x_k$ and $\xi_k$
are determined mod $4p_k$ by
\be
\ba 
x_k  &=q_k^{-1}  && \text{mod} \;
4p_k \ , \\ 
 \xi_k & = (p_k + q_k)^{-2} q_k 
&& \text{mod} \; 
4p_k \ .
\ea 
\ee 
For $N=2$ there is only one
 choice of $\mathbb Z_2$ grading on $\cC^k$:  
 $g(\lambda) = \phi(\lambda)$ mod 2.

Let us now discuss the case of $N$ an odd prime. The value of $\qq_k$ is 
$\qq_k = \mathfrak s_k^N$
with
\be \label{eq_qqk_Noddprime}
\mathfrak s^2_k = 
\left\{
\begin{array}{ll}
e^{2\pi i \frac{y_k}{p_k N}}
&
\text{if $N \!\!\not | \, q_k$} \ , \\
e^{2\pi i \frac{\eta_k}{p_k N}}
&
\text{if $N \, | \, q_k$} \ , 
\end{array}
\right.   
\ee
where $y_k$ is as in \eqref{eq_xy_def} while $\eta_k$ is
defined by
\be 
\eta_k =  (p_k+q_k)^{-2}  q_k \mod p_k N  \ .
\ee 
Moreover,
the grading function  is
\be \label{eq_v_choice_BIS}
g_k (\lambda) = 
\left\{ 
\begin{array}{ll}
q_k^{-1} \phi(\lambda)\; \text{mod} \; N  \ , & 
\text{if $N \!\!\not | \, q_k$} \ , 
\\
p_k^{-1} \phi(\lambda)\; \text{mod} \; N  \ , & 
\text{if $N \, | \, q_k$} \ . 
\\
\end{array}
\right.
\ee 
We notice that, since $N$ is prime and $\gcd(p_k,q_k)=1$,
if $N|q_k$ then $\gcd(N,p_k)=1$.

\noindent{\bf Match with 
CS-invariants of anyonic flat connections.}
We set $M_3 = [\{ (p_k,q_k)\}_{k=1}^n]$ throughout
this section.
Our definition
of $\mathsf{RFC} ( \mathfrak{sl}(N) , M_3 )$
implies the following
properties, which relate this RFC
to flat connections on $M_3$:
\begin{itemize}
\item The simple objects of 
$\mathsf{RFC} ( \mathfrak{sl}(N) , M_3 )$ are in 1-to-1 correspondence
with (connected components of)
anyonic flat $SL(N,\mathbb C)$
connections on $M_3$.
\item
Simple objects are partitoned
into sectors of definite $\mathbb Z_N$ charge
$g$, and (connected components of) anyonic flat connections are partitioned into sectors with definite $\ell$, see \eqref{eq_rho}. The correspondence 
between simple objects and anyonic flat connections
preserves these decompositions, with   
possible values of $g$ in bijection
with possible values of $\ell$.

\item If the simple object
$\boldsymbol \lambda$ corresponds to the
(connected component of the)
anyonic flat $SL(N,\mathbb C)$
connection $\rho$, the twist of $\boldsymbol{\lambda}$
matches the CS invariant of $\rho$,
\be \label{eq_theta_and_CS}
\theta_{\boldsymbol{\lambda}} = K e^{- 2 \pi i {\rm CS}(\rho)} \ , 
\ee 
where the constant $K$ depends
of $N$ and the Seifert data,
but not on $\boldsymbol{\lambda}$.
\end{itemize}
A proof of these properties is
given in appendix
\ref{app_CS_match}.
We discuss here some key aspects of
the derivation.

To start with, we consider a generic $N\ge 2$ and 
 impose the restriction
$\gcd(N,q_k)=1$
on all Seifert fibers.
We want to define a 1-to-1 map
between 
the set
${\mathcal R}^N_{\{ (p_k,q_k)\}_{k=1}^n}$
in \eqref{eq_R_set}
(anyonic flat connections)
and the set
${\mathcal S}^N_{\{ (p_k,q_k)\}_{k=1}^n}$
in \eqref{eq_def_cS}
(simple objects in the RFC).
We map $\nu_k^{(I)}$
to the following
$\lambda_k^{(I)}$,
\be \label{eq_map_to_lambda}
\lambda_k^{(I)} = \nu_k^{(I+1)} - \nu_k^{(I)} -1 \ , \quad 
I=1,\dots,N-1 \ . 
\ee 
One verifies that this map
defines a 1-to-1
correspondence between
${\mathcal R}^N_{\{ (p_k,q_k)\}_{k=1}^n}$
and ${\mathcal S}^N_{\{ (p_k,q_k)\}_{k=1}^n}$.
Here, we need to use 
the prescription for the grading function (\ref{eq_grading})
as well as the  assumption
$\gcd(N,q_k)=1$.
See appendix \ref{app_CS_match}
for more details.

As far as property
\eqref{eq_theta_and_CS} is concerned,
it is established by 
direct computation 
using 
\eqref{eq_rho},
\eqref{eq_rho_x_diag},
\eqref{eq_a_param_with_m},
\eqref{eq:CSformula},
\eqref{eq_twist},
\eqref{eq_combine_twists},
and
\eqref{eq_map_to_lambda}, working
separately in each Seifert fiber, as explained in
appendix 
\ref{app_CS_match}.
Here we simply point out that
our assignent 
\eqref{eq_qqk} for $\qq_k$
in terms of $N$, $p_k$,
$q_k$ is precisely engineered in such a way that 
\eqref{eq_theta_and_CS} holds.

Next, let us now consider the case of $N$ prime and relax the requirement
$\gcd(N,q_k)=1$.
Crucially, the map
\eqref{eq_map_to_lambda} from
$\nu_k^{(I)}$ to 
$\lambda_k^{(I)}$ needs to be modified in this case
(cfr.~\cite{Cui:2021lyi} for
$N=2$).
As a preliminary, we define
\be 
\nu_k^{(0)} := \nu_k^{(N)} - p_k \ . 
\ee 
Now we generalize \eqref{eq_map_to_lambda} by setting
\be  \label{eq_map_to_lambda_gen}
\lambda_k^{(I)} = 
\left\{ 
\begin{array}{ll}
\nu_k^{(I+1)}
- \nu_k^{(I)} -1 & \text{if $\gcd(N,q_k)=1$} \ , 
\\
\nu_k^{(\sigma[\ell](I)+1)}
- \nu_k^{(\sigma[\ell](I))} -1 & \text{if $\gcd(N,q_k)\neq 1$} \ .
\end{array}
\right.
\ee 
In the second line,
$\sigma[\ell]$ is the cyclic permutation
of $(0,1,2,\dots,N-1)$
that sends 0 to $\ell$,
i.e.~$\sigma[\ell](I) =I+\ell$ mod $N$.
In appendix \ref{app_CS_match}
we demonstrate that this is the correct definition 
in order to have flat connections with definite $\ell$ to be mapped to weights $\lambda_k$
with a definite $\mathbb Z_N$ charge.
For $N$ an odd prime,
it is essential here to use the prescription
\eqref{eq_v_choice_BIS} for the $\mathbb Z_N$ grading function. 

The property
\eqref{eq_theta_and_CS} 
also holds in the case of $N$ prime with $\gcd(N,q_k)=1$ relaxed.
As before, it is  established working fiberwise. For those fibers with $\gcd(N,q_k)=1$
there is nothing new to prove. For those fibers with
$\gcd(N,q_k) \neq 1$,
one uses 
\eqref{eq_rho},
\eqref{eq_rho_x_diag},
\eqref{eq_a_param_with_m},
\eqref{eq:CSformula},
\eqref{eq_twist},
and the new map
\eqref{eq_map_to_lambda_gen}.
We emphasise that
the assignments
\eqref{eq_qqk_N2},
\eqref{eq_qqk_Noddprime}
for $\qq_k$
in the $N=2$ and $N$ odd prime cases
are precisely engineered
in such a way that 
\eqref{eq_theta_and_CS} holds.
We refer the reader to appendix \ref{app_CS_match}
for more details.\\

\noindent{\bf Observation on fibers with $p_k=N$.}
We have the

\newtheorem{prop}{Proposition}
\begin{prop}
Fix $N\ge 2$ and consider the Seifert manifolds 
\begin{align}
M_3 & = [(p_1, q_1), \dots, (p_t,q_t), (N, q_{t+1}), \dots, (N, q_n)] \ , \nonumber 
\\
\widetilde M_3 & = [(p_1, q_1), \dots, (p_t,q_t), (N, 1)
] \ ,
\end{align}
where $\gcd(N,q_k)=1$ for $k=1,\dots,n$,
and $1\le t <n$, with
$p_k>N$ for $k=1,\dots,t$.
Then 
$\mathsf{RFC} ( \mathfrak{sl}(N) , M_3 )$
and 
$\mathsf{RFC} ( \mathfrak{sl}(N) , \widetilde M_3 )$
have the same modular data.
\end{prop}

Indeed, if the $k$th fiber has $p_k=N$,
then $ \Delta_{N,p_k}$
consists of a single element $\lambda_k=0$.
No matter the value of $q_k$,
this element is assigned
grading $g=0$.
Also, this fiber contributes
$\theta_{\lambda = 0}=1$
and $S_{\lambda=0,\mu=0}=1$
irrespective of the value of
$\qq_k$. In the graded
$\boxtimes$ construction
we effectively project
onto the $g=0$ sector in each fiber. We see that, as soon
as we have one fiber $(p_k=N, q_k)$, we can add any number of other fibers of the same kind,
we arbitrary $q$'s, without changing the T- and 
S-matrices.

In light of the above proposition,
there is no loss in generality
in restricting to Seifert data
in which at most one of the $p_k$
equals $N$, and setting $q_k=1$
for that fiber.\\

\begin{figure}
    \centering
\begin{tikzpicture}[thick,scale=0.9, every node/.style={scale=0.6}]
	\begin{pgfonlayer}{nodelayer}
		\node [style=bulk, fill = blue, label=below:$0$] (0) at (-6, 0) {};
		\node [style=bulk, fill = green, label=below:$1$] (1) at (-5, 0) {};
	\end{pgfonlayer}
	\begin{pgfonlayer}{edgelayer}
		\draw [style=lattice line] (0) to (1);
	\end{pgfonlayer}
\end{tikzpicture}
\hspace{0.5em}
\begin{tikzpicture}[thick,scale=0.9, every node/.style={scale=0.6}]
	\begin{pgfonlayer}{nodelayer}
		\node [style=bulk, fill = blue, label=below:$0$] (0) at (-6, 0) {};
		\node [style=bulk, fill = green, label=below:$1$] (1) at (-5, 0) {};
	\end{pgfonlayer}
	\begin{pgfonlayer}{edgelayer}
		\draw [style=lattice line] (0) to (1);
	\end{pgfonlayer}
\end{tikzpicture}
\hspace{0.5em}
\begin{tikzpicture}[thick,scale=0.9, every node/.style={scale=0.6}]
	\begin{pgfonlayer}{nodelayer}
		\node [style=bulk, fill = blue, label=below:$0$] (0) at (-6, 0) {};
		\node [style=bulk, fill = green, label=below:$1$] (1) at (-5, 0) {};
	\end{pgfonlayer}
	\begin{pgfonlayer}{edgelayer}
		\draw [style=lattice line] (0) to (1);
	\end{pgfonlayer}
\end{tikzpicture}
\hspace{0.5em}
\begin{tikzpicture}[thick,scale=0.9, every node/.style={scale=0.6}]
	\begin{pgfonlayer}{nodelayer}
		\node [style=bulk, fill = blue, label=below:$0$] (0) at (-6, 0) {};
		\node [style=bulk,fill = green, label=below:$1$] (1) at (-5, 0) {};
		\node [style=bulk, fill = blue, label=below:$2$] (2) at (-4, 0) {};
	\end{pgfonlayer}
	\begin{pgfonlayer}{edgelayer}
		\draw [style=lattice line] (0) to (2);
	\end{pgfonlayer}
\end{tikzpicture}
\caption{Fundamental Weyl alcove for $\mathfrak{sl}(2)$ with  $p=3, 3,3,4$ respectively. $\Z_2$-grading is depicted by the color blue (even) and green (odd).}
\label{fig:example3334}
\end{figure}
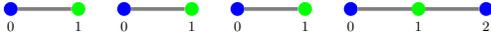

\noindent{\bf Example.} 
To illustrate this we now work out a concrete example, in preparation for the comprehensive list of models up to rank 5 in the next section. 
Consider the Seifert manifold with data
\be\label{OurM3}
M_3 = [\{(3,1), (3,1), (3,1), (4,1)\} ]\,,
\ee
and consider $G_{\C}= SL(2, \C)$. Here $\cZ_G= \Z_2$.  
We claim that this constructs the MTC of rank 3, given by (using the nomenclature of \cite{Ng:2023wsc}, this is model 6.~at rank $r=3$ in appendix E of that paper)
\be
 3_{\frac{15}{2},4.}^{16,639} \,.
 \ee
 This model has spins and S-matrix given by 
 \be
\ba
 h&=\left\{0,\frac{1}{2},\frac{15}{16}\right\}  \,, \cr 
 S&= \left(
\begin{array}{ccc}
 1 & 1 & \sqrt{2} \\
 1 & 1 & -\sqrt{2} \\
 \sqrt{2} & -\sqrt{2} & 0 \\
\end{array}
\right) \,.
\ea\ee
We obtain this MTC as the graded Deligne product
$\cC^1 \boxtimes_{\mathbb Z_2} 
\cC^2  \boxtimes_{\mathbb Z_2} 
\cC^3  \boxtimes_{\mathbb Z_2}\cC^4$, where
\begin{align}
\cC^1 &\cong \cC^2 \cong \cC^3\cong  \cC(\mathfrak{sl}(2) , p=3, \mathfrak{q}) \ , \;\; \mathfrak{q}= \mathfrak{s}^2 \ , \;\; \mathfrak{s} =e^{2\pi i \frac{1}{12}} \ ,
\nonumber
\\
\cC^4 &\cong  \cC(\mathfrak{sl}(2) , p=4, 
\mathfrak q) , \;\;
\mathfrak q  = \mathfrak s^2 
 , \;\; 
 \mathfrak{s}=
e^{2\pi i \frac{1}{16}} \  ,
\end{align}
where we have made use of \eqref{eq_qqk}.
In figure \ref{fig:example3334} we depict 
the fundamental Weyl alcoves
associated to $\cC^1$, $\cC^2$,
$\cC^3$, $\cC^4$ together with their
$\mathbb Z_2$ gradings. They have simple objects labelled by $\lambda_k^{(1)} \in \{0,1 \}$ with $k = 1,2,3$ and $\lambda_4^{(1)} \in \{0,1,2\}$, which are the weights inside the Weyl alcove. 
The spins and S-matrix of
$\mathcal{C}(\mathfrak{sl}(2), p, \mathfrak{q})$  where $\mathfrak{q} = \mathfrak{s}^2$ as specified in (\ref{eq_qqk}), are given by the specialization of (\ref{SmatSchopi}) to be 
\be
S_{\lambda_1, \lambda_2 }=\frac{\mathfrak{q}^{(\lambda_1+1)(\lambda_2+1)}-\mathfrak{q}^{-(\lambda_1+1)(\lambda_2+1)}}{\mathfrak{q}-\mathfrak{q}^{-1}} \,,
\ee
and (recall $\theta_\lambda = e^{2\pi i h_\lambda}$)
\be
\theta_{\lambda}=\mathfrak{q}^{\lambda(\lambda+2) / 2}
= \mathfrak s^{\lambda(\lambda+2)} \, .
\ee
Therefore for $\cC^1$ they are given by 
\be 
\cC^1 \;: \quad 
h = \left\{ 0 , \frac 14  \right \} \ , \quad 
S = \begin{pmatrix}
  1 &  1\\
  1 & -1
\end{pmatrix} \ . 
\ee
Those of $\cC^2$, $\cC^3$
are the same, while those of
$\cC^4$ read
\be 
\cC^4 \;: \quad 
h = \left \{ 0, \frac{3}{16} ,\frac 12 \right \} \ , \quad 
S = \begin{pmatrix}
1 & \sqrt{2} & 1 \\
\sqrt{2} & 0 &- \sqrt{2} \\
1 & -\sqrt{2} & 1
\end{pmatrix} \ . 
\ee

The resulting MTC $\cC^1 \boxtimes_{\mathbb Z_2} \dots \boxtimes_{\mathbb Z_2} \cC^4$ is again $\Z_2$-graded,
with simple objects labeled
by a tuple $\boldsymbol {\lambda} = (\lambda_1^{(1)}, \lambda_2^{(1)},\lambda_3^{(1)}, \lambda_4^{(1)})$.
More precisely:
\begin{align}
&\text{sector with $\mathbb Z_2$-charge $g=0$:} & 
& (0,0,0,0), \; (0,0,0,2) \ , 
\nonumber
\\ 
&\text{sector with $\mathbb Z_2$-charge $g=1$:} & 
& (1,1,1,1) \ . 
\end{align}
The spins are obtained by adding the spins in each of these sectors (mod 1), and the S-matrices by applying the graded Deligne product.

Note that $H_1(M_3, \Z_2)=0$ for (\ref{OurM3}), and the model is indeed modular. Note that the same rank 3 model in \cite{Cho:2020ljj} is realized with a Seifert manifold that has non-trivial $H_1(M_3, \Z_2)$, requiring them to gauge the $\Z_2$ 1-form symmetry, and rendering the model modular thereby.\\

\section{All low rank MTCs from \texorpdfstring{$\text{MTC}(M_3, G)$}{MTC(g,M3)}}
\label{sec:AllMTC}
Given {the} correspondence between Seifert manifolds and MTCs {described above}, it is natural to ask whether this provides a comprehensive list of MTCs. We do not have an answer to this question, but very strong evidence confirmed by low rank studies up to rank 5, that this may provide a complete classification. 

In appendix \ref{app:MTCTables} we provide a list of all MTCs from rank $r=1, \cdots, 5$ and a realization in terms of Seifert data of an $M_3$ data and a choice of $G_{ADE}$. Note that this is not unique, but the point here is that each MTC has such a realization. We compare this with the recent classification in  \cite{Ng:2023wsc}, and provide realizations for {\bf unitary and non-unitary} MTCs. The comparison is done to appendix E of  \cite{Ng:2023wsc}. The tables \ref{tab:r2}, \ref{tab:r3}, \ref{tab:r4}, \ref{tab:Rank5} provide the list of models for rank including $r=5$. 
Note that many models will be reducible, as in be a Deligne product of lower rank MTCs, this is indicated by ``(factors)". We provide a realization for all {\bf prime models}, in particular. 

We also explored rank 6 and in that case we find almost all prime models, and these are listed in appendix \ref{app:r6}. It would be interesting to extend the scan, and find realizations for all rank 6 models, and in fact complete the comparison to \cite{Ng:2023wsc} until rank 11. Note that this would also provide clarity as to whether the models that do not have a CS-type realization (denoted ``exotic models") may have a realization in terms of Seifert data.



\onecolumngrid

\vspace{1cm}
\noindent
\textbf{Acknowledgements.}
We thank Fabio Apruzzi, Lakshya Bhardwaj, Cyril Closset, Tudor Dimofte, Dongmin Gang,  Chris Negron, Pavel Putrov, Rajath Radhakrishnan, Zhenghan Wang, Martin Weidner, Xiao-Gang Wen for discussions at various stages of this work. SSN thanks King's College London for hospitality during some of this work. 
The work of SSN and JW is supported by the UKRI Frontier Research Grant, underwriting the ERC Advanced Grant "Generalized Symmetries in Quantum Field Theory and Quantum Gravity” and the Simons Foundation Collaboration on ``Special Holonomy in Geometry, Analysis, and Physics", Award ID: 724073, Schafer-Nameki. 
The work of FB is supported by the Simons Collaboration Grant on Global Categorical Symmetries.


\twocolumngrid

\bibliographystyle{ytphys}
\bibliography{GenSym}

\providecommand{\href}[2]{#2}\begingroup\raggedright\begin{thebibliography}{10}

\bibitem{Moore:1988qv}
G.~W. Moore and N.~Seiberg, ``{Classical and Quantum Conformal Field Theory},''
  \href{http://dx.doi.org/10.1007/BF01238857}{{\em Commun. Math. Phys.}
  {\bfseries 123} (1989) 177}.

\bibitem{Kitaev:2005hzj}
A.~Kitaev, ``Anyons in an exactly solved model and beyond,''
  \href{http://www.sciencedirect.com/science/article/pii/S0003491605002381}{{\em
  Annals of Physics} {\bfseries 321} no.~1, (Jan., 2006) 2--111}.

\bibitem{Wen:2015qgn}
X.-G. Wen, ``A theory of 2+{{1D}} bosonic topological orders,''
  \href{http://arxiv.org/abs/1506.05768}{{\em National Science Review}
  {\bfseries 3} no.~1, (Mar., 2016) 68--106},
  \href{http://arxiv.org/abs/1506.05768}{{\ttfamily arxiv:1506.05768}}.

\bibitem{EGNO}
P.~Etingof, S.~Gelaki, D.~Nikshych, and V.~Ostrik,
  \href{http://dx.doi.org/10.1090/surv/205}{{\em Tensor categories}}, vol.~205
  of {\em Mathematical Surveys and Monographs}.
\newblock American Mathematical Society, Providence, RI, 2015.
\newblock \url{https://doi.org/10.1090/surv/205}.

\bibitem{Simon:2023hdq}
S.~H. Simon, {\em {Topological Quantum}}.
\newblock Oxford University Press, 9, 2023.

\bibitem{mignard2021modular}
M.~Mignard and P.~Schauenburg, ``Modular categories are not determined by their
  modular data,'' {\em Letters in Mathematical Physics} {\bfseries 111} no.~3,
  (2021) 60.

\bibitem{Note1}
Although the first such ambiguity occurs at a relatively high rank, 49.

\bibitem{Ng:2023wsc}
S.-H. Ng, E.~C. Rowell, and X.-G. Wen, ``{Classification of modular data up to
  rank 11},''
\newblock 8, 2023.
\newblock \href{http://arxiv.org/abs/2308.09670}{{\ttfamily arXiv:2308.09670
  [math.QA]}}.

\bibitem{ostrik2002fusion}
V.~Ostrik, ``Fusion categories of rank 2,'' 2002.

\bibitem{ostrik2005premodular}
V.~Ostrik, ``Pre-modular categories of rank 3,'' 2005.

\bibitem{Rowell:2007dge}
E.~Rowell, R.~Stong, and Z.~Wang, ``On classification of modular tensor
  categories,'' Nov., 2009.
\newblock \url{http://arxiv.org/abs/0712.1377}.

\bibitem{bruillardClassificationModularCategories2016}
P.~Bruillard, S.-H. Ng, E.~C. Rowell, and Z.~Wang, ``On classification of
  modular categories by rank,''
  \href{http://arxiv.org/abs/1507.05139}{{\ttfamily arxiv:1507.05139 [math]}}.
  \url{http://arxiv.org/abs/1507.05139}.

\bibitem{ngReconstructionModularData2022}
S.-H. Ng, E.~C. Rowell, Z.~Wang, and X.-G. Wen, ``Reconstruction of modular
  data from \${{SL}}\_2(\textbackslash mathbb\{\vphantom\}{{Z}}\vphantom\{\})\$
  representations,'' \href{http://arxiv.org/abs/2203.14829}{{\ttfamily
  arxiv:2203.14829 [cond-mat, physics:math-ph]}}.
  \url{http://arxiv.org/abs/2203.14829}.

\bibitem{Dimofte:2011ju}
T.~Dimofte, D.~Gaiotto, and S.~Gukov, ``Gauge {{Theories Labelled}} by
  {{Three-Manifolds}},'' \href{http://arxiv.org/abs/1108.4389}{{\em
  arXiv:1108.4389 [hep-th]} (Aug., 2011) 1--58},
  \href{http://arxiv.org/abs/1108.4389}{{\ttfamily arxiv:1108.4389 [hep-th]}}.

\bibitem{Cho:2020ljj}
G.~Y. Cho, D.~Gang, and H.-C. Kim, ``M-theoretic {{Genesis}} of {{Topological
  Phases}},'' \href{http://arxiv.org/abs/2007.01532}{{\em Journal of High
  Energy Physics} {\bfseries 2020} no.~11, (Nov., 2020) 115},
  \href{http://arxiv.org/abs/2007.01532}{{\ttfamily arxiv:2007.01532}}.

\bibitem{Cui:2021lyi}
S.~X. Cui, Y.~Qiu, and Z.~Wang, ``{From Three Dimensional Manifolds to Modular
  Tensor Categories},''
  \href{http://dx.doi.org/10.1007/s00220-022-04517-4}{{\em Commun. Math. Phys.}
  {\bfseries 397} no.~3, (2023) 1191--1235},
  \href{http://arxiv.org/abs/2101.01674}{{\ttfamily arXiv:2101.01674
  [math.QA]}}.

\bibitem{Cui:2021yes}
S.~X. Cui, P.~Gustafson, Y.~Qiu, and Q.~Zhang, ``From torus bundles to
  particle\textendash hole equivariantization,''
  \href{http://undefined/article/10.1007/s11005-022-01508-3}{{\em Letters in
  Mathematical Physics} {\bfseries 112} no.~1, (Feb., 2022) 1--19}.

\bibitem{OrlikBook}
P.~Orlik, {\em Seifert Manifolds}.
\newblock Springer, Berlin, Heidelberg, 1972.

\bibitem{Neumann1978}
W.~D. Neumann and F.~Raymond, {\em Seifert manifolds, plumbing,
  {$mu$}-invariant and orientation reversing maps},
  \href{http://dx.doi.org/10.1007/BFb0061699}{pp.~163--196}.
\newblock Springer Berlin Heidelberg, Berlin, Heidelberg, 1978.
\newblock \url{https://doi.org/10.1007/BFb0061699}.

\bibitem{FuturePaper}
F.~Bonetti, S.~Schafer-Nameki, and J.~Wu, ``{To appear},''
\newblock 2024.

\bibitem{nishi}
H.~Nishi, ``Su(n)-chern–simons invariants of seifert fibered 3-manifolds,''
  {\em International Journal of Mathematics} {\bfseries 09} (1998) 295--330.

\bibitem{mignardModularCategoriesAre2021}
M.~Mignard and P.~Schauenburg,
  \href{http://dx.doi.org/10.48550/arXiv.1708.02796}{``Modular categories are
  not determined by their modular data,''} pp.~1--11.
\newblock 2021.
\newblock \href{http://arxiv.org/abs/1708.02796}{{\ttfamily 1708.02796}}.
\newblock \url{http://arxiv.org/abs/1708.02796}.

\bibitem{chariGuideQuantumGroups1995}
V.~Chari and A.~N. Pressley, {\em A {{Guide}} to {{Quantum Groups}}}.
\newblock {Cambridge University Press}, July, 1995.

\bibitem{sawinQuantumGroupsRoots2003}
S.~F. Sawin, ``Quantum {{Groups}} at {{Roots}} of {{Unity}} and
  {{Modularity}},'' \href{https://arxiv.org/abs/math/0308281v2}{{\em arXiv.org}
  (Aug., 2003) 1--30}.

\bibitem{Rowell:2005hv}
E.~C. Rowell, ``From {{Quantum Groups}} to {{Unitary Modular Tensor
  Categories}},'' Mar., 2006.
\newblock \url{http://arxiv.org/abs/math/0503226}.

\bibitem{schopierayLieTheoryFusion2018}
A.~Schopieray, ``Lie {{Theory}} for {{Fusion Categories}}: A {{Research
  Primer}},'' Oct., 2018.
\newblock \url{http://arxiv.org/abs/1810.09055}.

\bibitem{DiFrancesco:1997nk}
P.~Di~Francesco, P.~Mathieu, and D.~Senechal,
  \href{http://dx.doi.org/10.1007/978-1-4612-2256-9}{{\em {Conformal Field
  Theory}}}.
\newblock Graduate Texts in Contemporary Physics. Springer-Verlag, New York,
  1997.

\bibitem{kauffmanTemperleyLiebRecouplingTheory1994}
L.~H. Kauffman, S.~L. Lins, and S.~Lins, {\em Temperley-{{Lieb Recoupling
  Theory}} and {{Invariants}} of 3-Manifolds}.
\newblock {Princeton University Press}, July, 1994.

\bibitem{bruguieres2000categories}
A.~Bruguieres, ``Cat{\'e}gories pr{\'e}modulaires, modularisations et
  invariants des vari{\'e}t{\'e}s de dimension 3,'' {\em Mathematische Annalen}
  {\bfseries 316} no.~2, (2000) 215--236.

\bibitem{YellowBook}
P.~Francesco, P.~Mathieu, and D.~S{\'e}n{\'e}chal, {\em Conformal field
  theory}.
\newblock Springer Science \& Business Media, 2012.

\end{thebibliography}\endgroup

\onecolumngrid

\appendix

\section{More on Seifert Manifolds}
\label{sec:moreSeifert}

 We denote the Seifert 3-manifold
$M_3$ associated to
the Seifert data 
$b$, $g$, $\{ (p_k,q_k) \}_{k=1}^{n}$ as
\be 
M_3 = [b;g;\{ (p_k,q_k) \}_{k=1}^{n}] \ . 
\ee 
Orientation reversal is implemented
on Seifert data as $b \mapsto -b$,
$q_k \mapsto -q_k$.
If two
Seifert data $b$, $g$, $\{ (p_k,q_k) \}_{k=1}^{n}$
and 
$b'$, $g$, $\{ (p_k,q'_k) \}_{k=1}^{n}$
are related as
\be 
b' = b- 
\sum_{k=1}^n m_k  \ , \,\,  
q'_k = q_k +p_k m_k \ , \, \,
k=1,\dots,n \ , \label{eq:Seifertgauge}
\ee
with $m_k$ integers,
the associated 3-manifolds
are diffeomorphic (via an
orientation-preserving map).

Note that the simplifying assumption $\gcd(N,q_i)=1$ is not preserved under the Seifert `gauge' transformation \eqref{eq:Seifertgauge}, while keeping $b = 0$. Given a three manifold with Seifert data $M_3 = [0;0;\{ (p_k,q_k) \}_{k=1}^{n}]$, if certain $q_i$ is not coprime to $N$, we can often find a `gauge' where $\gcd(N,q_i)=1$ for all $i$ by means of \eqref{eq:Seifertgauge}. There are however cases where such `gauge' does not exist. The simplest example is $\left(\{(3,1),(3,1),(3,2) \}\right)$ for $N=2$ and it is easy to see that it is impossible to choose all three $q_i'$ to be odd by using  \eqref{eq:Seifertgauge}. 

Let's start with $N=2$ and any $n\geq 3$. Suppose two of the fibers have the Seifert data $\left\{(\mathrm{even},\mathrm{odd}),(\mathrm{even},\mathrm{odd})\right\}$, the parity will always be preserved under \eqref{eq:Seifertgauge}. Similarly for $\left\{(\mathrm{odd},\mathrm{odd}),(\mathrm{odd},\mathrm{even})\right\}$.
On the other hand, $\left\{(\mathrm{odd},\mathrm{even}),(\mathrm{odd},\mathrm{even})\right\}$ can be changed to $\left\{(\mathrm{odd},\mathrm{odd}),(\mathrm{odd},\mathrm{odd})\right\}$ and $\left\{(\mathrm{even},\mathrm{odd}),(\mathrm{odd},\mathrm{even})\right\}$ can be changed to $\left\{(\mathrm{odd},\mathrm{odd}),(\mathrm{even},\mathrm{odd})\right\}$. Based on this observation, for any $n\geq 3$ fibers, we can always reduce it to the following $n+2$ types.

We can have
\begin{equation}
    \left\{(\mathrm{odd},\mathrm{odd}),\dots, (\mathrm{odd},\mathrm{odd}), (\mathrm{even},\mathrm{odd}), \dots, (\mathrm{even},\mathrm{odd})\right\} \,,
\end{equation}
where the number of $(\mathrm{even},\mathrm{odd})$ fiber can be any number from $0$ to $n$. In addition, we have
\begin{equation}
    \left\{(\mathrm{odd},\mathrm{odd}),\dots, (\mathrm{odd},\mathrm{odd}), (\mathrm{odd},\mathrm{even}) \right\} \,,
\end{equation}
which is only case that has $\gcd(2,q_i) \neq 0$. Furthermore it has trivial $H_1(M_3,\mathbb{Z}_2)$ if and only if 
\begin{equation}
    KL = \left( \prod p_i\right) \left|\sum_i \frac{q_i}{p_i} \right|
\end{equation}
is odd, namely if and only if $n$ is even.

For $N=3$ and any $n\geq 3$, it is easy to see that one can always find $q_i'$ with $\gcd(N,q_i') =1$. So we don't lose any generality by assuming $\gcd(N,q_i) =1$ for all $i$.

For $N=4$, we are interested in whether it is possible to choose all $q_i'$ to be odd. Let $r_i$ be any integer with remainder $i$ divided by $4$, i.e. $r_i = i \mod 4$, for $i = 0,1,2,3$. If there are two fibers with the Seifert data $\left\{(\mathrm{odd},r_0),(\mathrm{odd},r_0)\right\}$, one can always go to a gauge where both $q$'s are odd. Similarly for $\left\{(\mathrm{odd},r_0),(\mathrm{odd},r_2)\right\}$ and $\left\{(\mathrm{odd},r_2),(\mathrm{odd},r_2)\right\}$. Therefore it suffices to consider the Seifert data with at most one $q_i$ being $r_0$ or $r_2$. In addition, any pair $\left\{(\mathrm{odd}, \mathrm{odd}),(\mathrm{odd},r_2)\right\}$ can be made into $\left\{(\mathrm{odd},\mathrm{odd}),(\mathrm{odd},r_0)\right\}$ and any pair $\left\{(\mathrm{even}, \mathrm{odd}),(\mathrm{odd},r_0)\right\}$ can be made into $\left\{(\mathrm{even}, \mathrm{odd}),(\mathrm{odd},\mathrm{odd})\right\}$. Therefore the only case with $\gcd(q_i,4)\neq 1$ that can't be removed by using  \eqref{eq:Seifertgauge}~is
\begin{equation}
    \left\{(\mathrm{odd},\mathrm{odd}),\dots, (\mathrm{odd},\mathrm{odd}), (\mathrm{odd},r_0) \right\}\,,
\end{equation}
where again $r_0$ is any multiple of $4$. We leave the discussion of general $N$ to future work.

\section{Mapping Anyons to Anyonic Flat Connections}
\label{app_CS_match}

In this appendix 
we verify the 1-to-1
correspondence between
simple objects of
$\mathsf{RFC}(\mathfrak{sl}(N), M_3)$
and anyonic $\mathfrak{sl}(N)$
flat connections on $M_3$.
Thoroughout this appendix
$M_3$ is the Seifert manifold
$M_3 = [\{(p_k,q_k) \}_{k=1}^n]$.

We distinguish two cases:
(1) $N \ge 2$ and $\gcd(N,q_k)=1$
for all $k=1,\dots,n$; (2)
$N$ prime but without any restriction on $\gcd(N,q_k)$. The most general case without any restrictions on $N$ and $\gcd(N,q_k)$ will be left for future work.

\subsection{The case where \texorpdfstring{$\gcd(N,q_k)=1$}{gcd} for all Seifert fibers}

Recall that a connected component in the moduli space of  anyonic flat connections
is determined by the data $(\ell, \nu_k^{(I)})$, $I=1,\dots,N$, satisfying 
$\sum_{I=1}^N \nu_k^{(I)} = \ell q_k$,
see 
\eqref{eq_rho},
\eqref{eq_rho_x_diag},
\eqref{eq_a_param_with_m},
\eqref{eq_R_set}.

Our first task is to 
exhibit a 1-to-1 correspondence
between 
simple objects in $\mathsf{RFC}(\mathfrak{sl}(N), M_3)$
and
anyonic $SL(N, \bC)$ flat connections on
$M_3$.
To this end,
we use the map from
flat connection data
$\nu_k^{(I)}$
to weight vectors $\lambda_k^{(I)}$
in \eqref{eq_map_to_lambda},
repeated here for convenience,
\be \label{eq_map_to_lambda_again}
\lambda_k^{(I)} = \nu_k^{(I+1)} - \nu_k^{(I)} -1 \ , \quad 
I=1,\dots,N-1 \ . 
\ee 
First, it is easy to verify that
$\lambda_k^{(I)} \in {\Delta}_{N,p_k}$,
for each $k=1,\dots,n$.
Next, we show that the $\mathbb Z_N$
charge $g$ of $\lambda_k^{(I)}$,
defined as 
$g = g_k(\lambda_k) = q_k^{-1} \phi(\lambda_k)$ mod $N$,
is independent of $k$ and determined
by $\ell$. 
Indeed, using the definition
of $\phi$ \eqref{eq_def_phi},
the map \eqref{eq_map_to_lambda_again},
and the property
$\sum_{I=1}^N \nu_k^{(I)}=\ell q_k$,
one computes
\be 
g = q_k^{-1 }\phi(\lambda_k)
= q_k^{-1 } \Big( - q_k \ell - \tfrac{N(N-1)}{2}\Big)
= - \ell - q_k^{-1 }\tfrac{N(N-1)}{2}
\mod N \ . 
\ee 
To proceed, we prove the following relation,
\be \label{eq_lemma}
q_k^{-1 }\tfrac{N(N-1)}{2}=
\tfrac{N(N-1)}{2} \mod N  \ . 
\ee 
If $N$ is odd, $(N-1)/2$ is integer and both sides vanish mod $N$.
If $N$ is even, $q_k$ is odd because
$\gcd(N,q_k)=1$. Hence $q_k^{-1}$ is odd and $q_k^{-1} -1$ is even.
As a result $(q_k^{-1} -1) \tfrac{N(N-1)}{2}=0$ mod $N$, giving \eqref{eq_lemma}. In summary,
we have
\be \label{eq_g_and_ell}
g = -\ell - \tfrac{N(N-1)}{2} \mod N \ , 
\ee
independent of $k$ and determined by $\ell$, as claimed. Using 
\eqref{eq_g_and_ell} it is straightforward to verify that
\eqref{eq_map_to_lambda_again} defines
a 1-to-1 map
from ${\mathcal R}^N_{\{ (p_k,q_k)\}_{k=1}^n}$
to $\mathcal S^N_{\{ (p_k,q_k)\}_{k=1}^n}$.

Next,
we want to establish the match between
CS invariants of anyonic flat connections and twists in the RFC.
More precisely, let us start with an anyonic flat connection with data
$(\ell, \nu_k^{(I)})$. Its CS invariant
is given as a sum of terms,
one for each Seifert fiber,
see \eqref{eq:CSformula},
\be 
{\rm CS} = \sum_{k=1}^n {\rm CS}_k 
\mod \mathbb Z
\ , \qquad 
{\rm CS}_k := \frac 12 p_k r_k {\rm Tr} X_k^2
- \frac 12 q_k s_k {\rm Tr}H^2 \mod \mathbb Z \ . 
\ee 
Using 
\eqref{eq_rho},
\eqref{eq_rho_x_diag},
\eqref{eq_a_param_with_m},
\eqref{eq:CSformula}, and
\eqref{eq_def_X_and_H}, we compute
\be \label{eq_CSk}
{\rm CS}_k
=  \frac{r_k}{2p_k} \sum_{I=1}^N (\nu_k^{(I)})^2
- \frac{\ell^2 q_k^2}{2 N p_k}
+ \frac{(N-1) \ell^2}{N} \mod \mathbb Z \ , 
\ee 
where we have made use of $\sum_{I=1}^N \nu_k^{(I)}  = \ell q_k$.
We want to compare $e^{ 2\pi i {\rm CS}_k}$
with the
the quantity
\be 
 \theta_{\lambda_k}= \qq_k^{\langle \lambda_k , \lambda_k + 2\rho \rangle } \ ,
\ee 
see \eqref{eq_twist},
which is the 
contribution of the $k$th
factor in the graded $\boxtimes$ presentation of the RFC to the total twist of the simple object
labeled by $\boldsymbol{\lambda}= (\lambda_k)_{k=1}^n$.
We write $\qq_k = e^{2\pi i \kappa_k/(2p_k N)}$ and therefore
\be  \label{eq_theta_massaged}
\theta_k = \exp 2\pi i \bigg[ 
\frac{\kappa_k}{2 p_k N}  
\sum_{I,J=1}^{N-1} B^{(N)}_{IJ}
\lambda_k^{(I)} \left( \lambda_k^{(J)} +2\right)
\bigg] \ .
\ee 
We have introduced the symmetric 
$(N-1) \times (N-1)$
matrix
$B^{(N)}_{IJ}$
that represents $N$ times the interior
product $\langle \cdot , \cdot \rangle$
on the weight lattice.
More explicitly, 
\be 
B^{(N)}_{IJ} = N (A^{(N)-1})_{IJ} = 
N \min(I,J) - IJ \ ,
\ee 
where $A^{(N)}$ is the Cartan
matrix of $\mathfrak{sl}(N)$.
We notice that the entries of $B$ are integers.
We are now in a position
to plug
\eqref{eq_map_to_lambda_again} into 
\eqref{eq_theta_massaged}, and compare
with
\eqref{eq_CSk}.
Our goal is to obtain a relation of the form
\be  \label{eq_goal}
\theta_{\lambda_k} = K_k e^{- 2\pi i {\rm CS}_k} \  , 
\ee 
where $K_k$ is a non-zero constant,
depending on $N$ and the Seifert data
$p_k$, $q_k$, but independent of $\ell$
and $\nu_k^{(I)}$.
We verify that \eqref{eq_goal}
can indeed be achieved
(independently for each $k$)
precisely by choosing 
$\kappa_k$ in such a way that
$\qq_k$ is given as in 
\eqref{eq_qqk} in terms of $N$
and the Seifert data $p_k$, $q_k$.

Having established 
\eqref{eq_goal}, by taking a product over $k=1,\dots,n$ we arrive
at the desired relation between 
the twist and the CS invariant,
\be 
\theta_{\boldsymbol{\lambda}} = K e^{- 2 \pi i {\rm CS}} \ , 
\ee 
where $K=\prod_{k=1}^n K_k$.

\subsection{The case \texorpdfstring{$N$}{N} prime with some Seifert fibers with 
\texorpdfstring{$\gcd(N,q_k)\neq 1$}{gcd}}

We now restrict $N$ to be prime,
but we allow Seifert fibers for which
$\gcd(N,q_k) \neq 1$.

Let us focus on a fiber with label $k$
for which $\gcd(N,q_k) \neq 1$.
Since $N$ is prime,
it follows that $N | q_k$.
Firstly, let us argue why
we must modify the map
\eqref{eq_map_to_lambda}
from eigenvalue data $(\ell, \nu_k^{(I)})$ to weights $\lambda_k^{(I)}$.
If we were to make use of \eqref{eq_map_to_lambda},
a direct computation shows that 
the $N$-ality function $\phi(\lambda)$
 in \eqref{eq_def_phi} can be evaluated in terms of the 
eigenvalue data as
\be 
\text{using \eqref{eq_map_to_lambda}:} \qquad 
\phi(\lambda_k) = 
- \ell q_k - \frac{N(N-1)}{2} + N\nu_k^{(N)}
=  \left\{ 
\begin{array}{lll}
- \ell q_k - 1 & \mod 2  & \text{if $N=2$} \ , 
\\
- \ell q_k  & \mod N  & \text{if $N$ is an odd prime}  \ . 
\end{array}
\right.
\ee 
Since $N|q_k$, we see that the function
$\phi(\lambda)$ assigns the same value
to \emph{all} flat connections,
irrespectively of their $\ell$.
On the other hand,
we know that the possible $\mathbb Z_N$
gradings on 
$\mathcal{C}(\mathfrak{sl}(N), p, \mathfrak{q})$
are all proportional to the function
$\phi(\lambda)$, see \eqref{eq_grading}.
It is thus impossible 
to define a grading
on $\mathcal{C}(\mathfrak{sl}(N), p, \mathfrak{q})$
such that 
weights $\lambda$ with
different gradings $g\in \mathbb Z_N$
correspond to anyonic
flat connections with different values of $\ell$.

Building on \cite{Cui:2021lyi},
we overcome this obstacle by 
modifying the map 
\eqref{eq_map_to_lambda}
from eigenvalue data $(\ell, \nu_k^{(I)})$ to weights $\lambda_k^{(I)}$.
We proceed as follows.
Define
\be 
\nu_k^{(0)} = \nu_k^{(N)} - p_k \ . 
\ee 
Let $\sigma[\ell]$ denote the
cyclic permutation of
$(0,1,2,\dots, N-1)$
that assigns the value $\ell$ to $0$,
more explicitly,
\be 
\sigma[\ell]  = \begin{pmatrix}
    0 & 1 & 2 & 3 & \dots & N-1 \\ 
    \ell & \ell+1 & \ell+2 & \ell+3 & \dots & \ell+N-1
\end{pmatrix} \ , 
\ee 
where all integers in the second row
are understood to be reduced mod $N$
to lie in the range $\{0,\dots, N-1\}$.
With this notation,
the proposed modification of
\eqref{eq_map_to_lambda} is as follows,
\be  \label{eq_map_to_lambda_generalized}
\lambda_k^{(I)} = 
\left\{ 
\begin{array}{ll}
\nu_k^{(I+1)}
- \nu_k^{(I)} -1 & \text{if $\gcd(N,q_k)=1$} \ , 
\\
\nu_k^{(\sigma[\ell](I)+1)}
- \nu_k^{(\sigma[\ell](I))} -1 & \text{if $\gcd(N,q_k)\neq 1$} \ .
\end{array}
\right.
\ee 
In other words, 
when $\gcd(N,q_k)\neq 1$
we use a different cyclic permutation
for each value of $\ell$.
This definition is motivated by the fact
that it ensures
\be 
\ba 
&\text{if $N=2$:} & 
\phi(\lambda_k) &= 
\left\{ 
\begin{array}{lll}
- \ell q_k -1 & \mod 2  \ , & \text{if $q_k$ is odd} \ , 
\\
- \ell p_k -1
& \mod 2  \ , & \text{if $q_k$ is even} \ ,
\end{array}
\right.
\quad 
\text{hence}
\quad 
\phi(\lambda_k) = \ell+1 \mod 2 \ . 
\\
&\text{if $N$ is an odd prime:} & 
\phi(\lambda_k) &= 
\left\{ 
\begin{array}{lll}
- \ell q_k & \mod N  \ , & \text{if $\gcd(N,q_k)=1$} \ , 
\\
- \ell p_k & \mod N  \ , & \text{if $\gcd(N,q_k)\neq 1$} \ .
\end{array}
\right.
\ea 
\ee 
For $N=2$, we see that we have achieved our goal of having a 1-to-1 correspondence between $\lambda$'s
with definite $\phi$ mod 2,
and anyonic flat connections with definite
$\ell$.
In the case of $N$ odd prime,
we can also achieve this goal
by choosing the 
$\mathbb Z_N$ grading function
as in \eqref{eq_v_choice_BIS},
\be 
\text{if $N$ is an odd prime:} \quad 
g_k(\lambda) = 
\left\{ 
\begin{array}{lll}
q_k^{-1} \phi(\lambda) \mod N  \ , & \text{if $\gcd(N,q_k)=1$} \ , 
\\
p_k^{-1}\phi(\lambda) \mod N  \ , & \text{if $\gcd(N,q_k)\neq 1$} \ ,
\end{array}
\right.
\quad 
\text{hence} \quad 
g_k(\lambda_k) = -\ell \mod N \ .  
\ee 

We see that our definitions
are engineered in such a way that 
\be 
g = g_k(\lambda_k) = -\ell - \tfrac{N(N-1)}{2} \mod N \ , 
\ee 
for every prime $N$ and for every fiber $k$ 
(including those fibers with
$\gcd(N,q_k)\neq 1$).
But this is exactly the same result as in the case of general $N$ with the constraint $\gcd(N,q_k)=1$,
see \eqref{eq_g_and_ell}.
By reasoning as in that previous case,
one shows that 
the map
\eqref{eq_map_to_lambda_generalized} defines a 1-to-1 correspondence between
anyonic flat connections and simple objects of the RFC.

Finally, we have to show that
the twist of simple objects in the RFC matches with the CS invariant of anyonic flat connections.
We proceed as in the previous
subsection.
It is convenient to 
analyze each Seifert fiber separately:
we have to prove
\eqref{eq_goal}
for each fiber.
For those fibers with
$\gcd(N,q_k)=1$ we can
recycle the proof of the previous subsection verbatim.
For those fibers with
$\gcd(N,q_k)\neq 1$,
the proof has to be revised, but the strategy is the same.
On the CS side,
we again make use of
\eqref{eq_rho},
\eqref{eq_rho_x_diag},
\eqref{eq_a_param_with_m},
\eqref{eq:CSformula}, and
\eqref{eq_def_X_and_H}.
On the twist side, the only modification compared to the previous case
is how we express $\lambda_k^{(I)}$
in terms of $\nu_k^{(I)}$:
we have to make use of
\eqref{eq_map_to_lambda_generalized}.
A direct computation shows that we can indeed achieve 
\eqref{eq_goal}
provided we fix $\qq_k$
according to the prescriptions
\eqref{eq_qqk_N2}
and 
\eqref{eq_qqk_Noddprime} in the main text, for $N=2$ and $N$ an odd prime, respectively.

\section{Fundamental Weyl Alcove for Simply-laced \texorpdfstring{$\mathfrak{g}$}{g}}
\label{app:Alcove}

Let $\mathfrak{g}$ be an ADE
Lie algebra with $\text{rank}(\mathfrak g) = r$.
An element $\lambda$ in the weight lattice is expanded onto fundamental weights
$\{ \omega_I\}_{I=1}^r$ as 
$\lambda = \sum_{I=1}^r \lambda^{(I)} \omega_I$,
where $\lambda^{(I)}$
are the Dynkin labels of $\lambda$.

The fundamental affine Weyl alcove  $ \Delta_{\mathfrak g, p}$
is defined as
\be 
 \Delta_{\mathfrak g, p} := \{ 0 < \langle \lambda + \rho , \alpha^\vee \rangle < p
\ , \alpha \in \Phi^+ \} \ ,
\ee 
where $\Phi^+$ is the set of positive roots, $\alpha^\vee =
\frac{2\alpha}{ \langle \alpha , \alpha \rangle}$
(which equals $\alpha$ for ADE algebras), and $\rho$ is the Weyl vector $\rho =\sum_{I=1}^r
\omega_I$.
We can equivalently write
\be 
 \Delta_{\mathfrak g,p} = \{ \langle \lambda + \rho , \alpha_I^\vee \rangle >0 \ , \; \;
\langle \lambda + \rho , \theta \rangle <p \} \ ,  
\ee 
where $\alpha_I$ are the simple roots of $\mathfrak g$
and $\theta$ is the highest root, 
\be 
\theta = \sum_{i=1}^r a_I  \alpha_I = \sum_{i=1}^r a_I^\vee  \alpha_I^\vee  \ , 
\ee 
where the expansion coefficients $a_I$, $a^\vee_I$ are the marks, comarks of $\mathfrak g$ ($a_I = a^\vee_I$ for simply laced algebras).
In terms of the Dynkin labels
$\lambda^{(I)}$ we can write more explicitly
\be \label{eq:generic_alcove}
 \Delta_{\mathfrak g,p} = \left \{ \lambda^{(I)} \geq 0 \ , \;\; \sum_{I=1}^r a_I^\vee \lambda^{(I)} \leq  p - h^\vee 
\right \} \ ,
\ee 
where we have introduced the dual Coxeter number $h^\vee$ of $\mathfrak g$,
\be 
h^\vee = 1 + \sum_{I=1}^r a_i^\vee \ . 
\ee 
From \eqref{eq:generic_alcove}, we see that  $ \Delta_{\mathfrak g,p}$ is precisely the affine Weyl alcove, i.e.~the set of integrable highest weight, for the affine algebra of $\mathfrak{g}$ at level $p-h^\vee$. Below we collect the comarks
and dual Coxeter numbers
for all ADE Lie algebras
(see e.g.~\cite{YellowBook}),
\be 
\ba 
A_r &: & 
\left(a_I^\vee\right)_{I=1}^r & = (1,1,1,\dots,1) \ ,  & 
h^\vee & = r+1 \\ 
D_r &: & 
\left(a_I^\vee\right)_{I=1}^r & = (1,2^{r-3} , 1,1 ) \ ,  & 
h^\vee & = 2r-2 \\ 
E_6 &: & 
\left(a_I^\vee\right)_{I=1}^r & = (1,2,3,2,1,2 ) \ ,  & 
h^\vee & = 12 \\ 
E_7 &: & 
\left(a_I^\vee\right)_{I=1}^r & = (  2,3,4,3,2,1,2 ) \ ,  & 
h^\vee & = 18 \\ 
E_8 &: & 
\left(a_I^\vee\right)_{I=1}^r & = (2,3,4,5,6 ,4,2,3 ) \ , 
 & 
h^\vee & = 30 \ .
\ea 
\ee 
where an exponent $n$ indicates
that an entry is repeated $n$ times.
We are using the standard ordering of simple roots,
corresponding to the standard form of the Cartan matrices for ADE Lie algebras
(as in \cite{YellowBook}).
For $\mathfrak g = A_{N-1}$,
\eqref{eq:generic_alcove} reduces to \eqref{eq_def_Delta}.

\section{Tables of MTCs from \texorpdfstring{$M_3$}{M3} for Rank \texorpdfstring{$r\leq 5$}{r<=5}
 }
\label{app:MTCTables}

We now identify the low rank MTCs, including non-unitary ones, with Seifert manifolds and a choice of $SL(N,\C)$. To label the low rank MTCs we use the notation in \cite{Ng:2023wsc}. In particular, each modular data in the list is labeled by $r_{c, D^2}^{\mathrm{ord}(T), \mathrm{fp}}$, where $r$ is the rank of the MTC, i.e.~the number of the simple objects, $c$ is the chiral central charge, $D^2 \equiv \sum_i d_i^2$ is the total quantum dimension.  $\operatorname{ord}(T)$ is the order of the T-matrix, i.e.~the smallest positive integer $n$ such that $T^n = \mathrm{id}$. $\mathrm{fp}$ is the ``finger print'', defined by the first three digits of $\left|\sum_i\left(s_i^2-\frac{1}{4}\right) d_i\right|$.

We list the rank and spins, or equivalently the T-matrix, but for all the models we tabulate we also checked agreement of the S-matrix with  \cite{Ng:2023wsc}. Labels are as in \cite{Ng:2023wsc} and $\#$ indicates the model number within the set of MTCs of that rank in \cite{Ng:2023wsc}.

\noindent{\bf Summary of results:} 
\begin{enumerate}
\item Rank 1: trivially realized. 
\item Rank 2: all models in \cite{Ng:2023wsc}. 
\item Rank 3: all models in \cite{Ng:2023wsc} including pseudo-unitary ones are realized, using $N=2$, $n=3$ and $n=4$, and for the final two models (number 1. and 2.) we need $N=3$ and $n=3$. 

\item Rank 4: all prime models in \cite{Ng:2023wsc}. 

\item Rank 5: all models in \cite{Ng:2023wsc}. 

\item Rank 6: See appendix \ref{app:r6}.
\end{enumerate}

Let us make a note on pseudo-unitary models.
A pseudo-unitary model has anyons of negative quantum dimension $d_i = S_{0i}/S_{00}$. It is related to a unitary model, as follows.
The T- and S-matrices of the unitary model are given in terms of those of the pseudo-unitary model by
\be 
e^{2\pi i \theta'_j}
= \sign(d_j) e^{2\pi i \theta_j} \ , \qquad 
S'_{ij} = \sign(d_i)
\sign(d_j) S_{ij} \ . 
\ee

\begin{table}[h!]
$$
\begin{array}{|c|c|c|c|c|c|c|}\hline 
N & n&  M_3 (\{(p_i, q_i)\}) & \text{Rank} 
& \text{Spins} & \text{Label} & \# \cr \hline \hline
2&3 & \{( 2,1), (3 ,1) ,(5 ,  1)\} & 2 & \{ 0, \frac{2}{5}   \} &   2_{\frac{14}{5},3.618}^{5,395} &3.  \text{(prime)} \cr \hline 
2&3 & \{( 2,1), (3 ,1) ,(5 ,  3)\} & 2 & \{ 0, \frac{4}{5}    \} &  2_{\frac{38}{5},1.381}^{5,491}   & 6.   \text{(prime)}  \cr \hline 
2&3 & \{( 2,1), (3 ,1) ,(5 ,  7)\} & 2 & \{  0, \frac{1}{5}   \} & 2_{\frac{2}{5},1.381}^{5,120}    &5.   \text{(prime)}  \cr \hline 
2&3 & \{( 2,1), (3 ,1) ,(5 ,  9)\} & 2 & \{  0, \frac{3}{5}   \} &  2_{\frac{26}{5},3.618}^{5,720}   & 4.  \text{(prime)} \cr \hline 
2&3 & \{( 3,1), ( 3,1) ,( 3,  1)\} & 2 & \{  0 , \frac{3}{4}  \} & 2_{7,2.}^{4,625}  &  2.  \text{(prime)} \sim 7.  \text{(prime, pseudo)}  \cr \hline 
2&3 & \{( 3,1), ( 3,1) ,( 3,  7)\} & 2 & \{ 0, \frac{1}{4}   \} &  2_{1,2.}^{4,437}  &  1.  \text{(prime)} \sim 8.  \text{(prime, pseudo)}  \cr \hline\hline 
2 &4  &  \{( 3,1), ( 3,1) , ( 3,1) ,( 3,  2)\}  & 2 & \{  0 , \frac{1}{4}  \} & 2_{1,2.}^{4,625}  &   7.  \text{(prime, pseudo)} \sim 2.  \text{(prime)} \cr \hline 
2 &4  &  \{( 3,1), ( 3,1) , ( 3,1) ,( 3,  4)\} &2 & \{ 0, \frac{3}{4}   \} &  2_{7,2.}^{4,562}  &   8.  \text{(prime, pseudo)} \sim 1.  \text{(prime)} \cr \hline\hline 

\end{array}
$$
\caption{MTCs from 3-manifolds: Rank $2$. Note models $7.$ and $8.$ are pseudo-unitary and related to 2. and 1., respectively.  \label{tab:r2}}
\end{table}


\begin{table}
$$
\begin{array}{|c|c|c|c|c|c|c|}\hline 
N & n&  M_3 (\{(p_i, q_i)\}) & \text{Rank} 
& \text{Spins} & \text{Label} & \# \cr \hline \hline
2&3 & \{( 2,1), ( 3,1), (7 ,1)\} & 3 & \left\{ 0,\frac{2}{7},\frac{6}{7}   \right\} &  3_{\frac{8}{7},9.295}^{7,245}  &  20. \text{(prime)}\cr \hline 
2&3 & \{( 2,1), ( 3,1), (7 ,3)\} & 3 &   \left\{0,\frac{2}{7},\frac{3}{7}\right\}    &  3_{\frac{12}{7},2.862}^{7,768}  &   21. \text{(prime, not-pseudo)} \cr \hline 
2&3 & \{( 2,1), ( 3,1), (7 ,5)\} & 3 &  \left\{0,\frac{4}{7},\frac{6}{7}\right\}     &  3_{\frac{52}{7},1.841}^{7,604}  &  24. \text{(prime, not-pseudo)} \cr \hline 
2&3 & \{( 2,1), ( 3,1), (7 ,9)\} & 3 & \left\{0,\frac{1}{7},\frac{3}{7}\right\}      &  3_{\frac{4}{7},1.841}^{7,953}  &  23. \text{(prime, not-pseudo)} \cr \hline 
2&3 & \{( 2,1), ( 3,1), (7 ,11)\} & 3 & \left\{0,\frac{4}{7},\frac{5}{7}\right\}      &  3_{\frac{44}{7},2.862}^{7,531}  &  22. \text{(prime, not-pseudo)} \cr \hline 
2&3 & \{( 2,1), ( 3,1), (7 ,13)\} & 3 & \left\{0,\frac{1}{7},\frac{5}{7}\right\}      & 3_{\frac{48}{7},9.295}^{7,790}  &  19. \text{(prime)} \cr \hline 
2&3 & \{( 3 ,1), ( 3 ,1), ( 4 ,1)\} & 3 &  \left\{0,\frac{1}{2},\frac{11}{16}\right\}    &  3_{\frac{11}{2},4.}^{16,648}  & 13. \text{(prime)}  \cr \hline 
2&3 & \{( 3 ,1), ( 3 ,1), ( 4 ,3)\} & 3 &  \left\{0,\frac{1}{2},\frac{9}{16}\right\}    &  3_{\frac{9}{2},4.}^{16,343}  &  17. \text{(pseudo)}\sim 3.  (\text{prime}) \cr \hline 
2&3 & \{( 3 ,1), ( 3 ,1), ( 4 ,5)\} & 3 &   \left\{0,\frac{1}{2},\frac{15}{16}\right\}   &  3_{\frac{15}{2},4.}^{16,113}  & 18. \text{(pseudo)}\sim 4. (\text{prime}) \cr \hline 
2&3 & \{( 3 ,1), ( 3 ,1), ( 4 ,7)\} & 3 &  \left\{0,\frac{1}{2},\frac{13}{16}\right\}  & 3_{\frac{13}{2},4.}^{16,330}   &  14. (\text{prime}) \cr \hline 
2&3 & \{( 3 ,1), ( 3 ,1), ( 4 ,9)\} & 3 &  \left\{0,\frac{3}{16},\frac{1}{2}\right\}    &   3_{\frac{3}{2},4.}^{16,553} &  11. (\text{prime})\cr \hline 
2&3 & \{( 3 ,1), ( 3 ,1), ( 4 ,11)\} & 3 &  \left\{0,\frac{1}{16},\frac{1}{2}\right\}    &  3_{\frac{1}{2},4.}^{16,980} & 15. \text{(pseudo)} \sim 5.  (\text{prime})  \cr \hline 
2&3 & \{( 3 ,1), ( 3 ,1), ( 4 ,13)\} & 3 &  \left\{0,\frac{7}{16},\frac{1}{2}\right\}    &  3_{\frac{7}{2},4.}^{16,167}  & 16.  \text{(pseudo)} \sim 6.  (\text{prime})  \cr \hline 
2&3 & \{( 3 ,1), ( 3 ,1), ( 4 ,15)\} & 3 &  \left\{0,\frac{5}{16},\frac{1}{2}\right\}    &  3_{\frac{5}{2},4.}^{16,465}  &  12.  \text{(prime)}  
\cr
\hline \hline 
2&4 & \{( 3 ,1), ( 3 ,1),( 3 ,1), ( 4 ,1)\} & 3 & \left\{0,\frac{1}{2},\frac{15}{16}\right\}   & 3_{\frac{15}{2},4.}^{16,639} & 6. (\text{prime})  \cr \hline 
2&4 & \{( 3 ,1), ( 3 ,1),( 3 ,1), ( 4 ,3)\} & 3 &  \left\{0,\frac{1}{2},\frac{13}{16}\right\}  & 3_{\frac{13}{2},4.}^{16,830} & 10. (\text{pseudo}) \sim 12.\text{(prime)} \cr \hline 
2&4 & \{( 3 ,1), ( 3 ,1),( 3 ,1), ( 4 ,5)\} & 3 & \left\{0,\frac{3}{16},\frac{1}{2}\right\}   &  3_{\frac{3}{2},4.}^{16,538}& 7. (\text{pseudo}) \sim 13. \text{(prime)}  \cr \hline 
2&4 & \{( 3 ,1), ( 3 ,1),( 3 ,1), ( 4 ,7)\} & 3 & \left\{0,\frac{1}{16},\frac{1}{2}\right\}   &  3_{\frac{1}{2},4.}^{16,598}&  3.  \text{(prime)} \cr \hline 
2&4 & \{( 3 ,1), ( 3 ,1),( 3 ,1), ( 4 ,9)\} & 3 &  \left\{0,\frac{7}{16},\frac{1}{2}\right\}  & 3_{\frac{7}{2},4.}^{16,332} & 4.  \text{(prime)}  \cr \hline 
2&4 & \{( 3 ,1), ( 3 ,1),( 3 ,1), ( 4 ,11)\} & 3 & \left\{0,\frac{5}{16},\frac{1}{2}\right\}   & 3_{\frac{5}{2},4.}^{16,345} &  8.  \text{(pseudo)} \sim 14.  \text{(prime)}\cr \hline 
2&4 & \{( 3 ,1), ( 3 ,1),( 3 ,1), ( 4 ,13)\} & 3 &  \left\{0,\frac{1}{2},\frac{11}{16}\right\}  & 3_{\frac{11}{2},4.}^{16,564} & 9.  \text{(pseudo)} \sim 11.  \text{(prime)} \cr \hline 
2&4 & \{( 3 ,1), ( 3 ,1),( 3 ,1), ( 4 ,15)\} & 3 & \left\{0,\frac{1}{2},\frac{9}{16}\right\}  &  3_{\frac{9}{2},4.}^{16,156}& 5. \text{(prime)} \cr \hline \hline
3& 3& \{(4,1), (4,1), (4,5)\} & 3  & \{0, {\frac13}, {\frac13}\} & 3_{2,3}^{3,527} & 1.  (\text{prime}) \cr \hline 
3& 3& \{(4,1), (4,5), (4,5)\} & 3  & \{0, {\frac23}, {\frac23}\} & 3_{6,3}^{3,138} & 2. (\text{prime}) \cr \hline \hline 
\end{array}
$$
\caption{MTCs from 3-manifolds: Rank 3.  The table shows the complete set of rank 3 models from $SL(2, \C)$ and $n=3, 4$-fibers: Rank $3$ and $SL(3,\C)$ with $n=3$ fibers. \label{tab:r3}}
\end{table}


\begin{table}
$$
\begin{array}{|c|c|c|c|c|c|c|}\hline 
N & n&  M_3 (\{(p_i, q_i)\}) & \text{Rank} 
& \text{Spins} & \text{Label} & \# \cr \hline \hline
2&3 & \{(2,1), ( 3,1), (9 ,1)\} & 4 &   \left\{0,\frac{2}{9},\frac{1}{3},\frac{2}{3}\right\} &4_{\frac{10}{3},19.23}^{9,459} & 24.  \text{(prime)}\cr \hline 
2&3 & \{(2,1), ( 3,1), (9 ,5)\} & 4 &  \left\{0,\frac{1}{3},\frac{4}{9},\frac{2}{3}\right\}  & 4_{\frac{14}{3},5.445}^{9,544} & 26.   \text{(prime)}\cr \hline 
2&3 & \{(2,1), ( 3,1), (9 ,7)\} & 4 &  \left\{0,\frac{1}{3},\frac{2}{3},\frac{8}{9}\right\}  & 4_{\frac{22}{3},2.319}^{9,549}&   29.\text{(prime)}\cr \hline 
2&3 & \{(2,1), ( 3,1), (9 ,11)\} & 4 &  \left\{0,\frac{1}{9},\frac{1}{3},\frac{2}{3}\right\} &  4_{\frac{2}{3},2.319}^{9,199} & 28.   \text{(prime)}\cr \hline 
2&3 & \{(2,1), ( 3,1), (9 ,13)\} & 4 &  \left\{0,\frac{1}{3},\frac{5}{9},\frac{2}{3}\right\}  &4_{\frac{10}{3},5.445}^{9,616}&  27. \text{(prime)}\cr \hline 
2&3 & \{(2,1), ( 3,1), (9 ,17)\} & 4 &  \left\{0,\frac{1}{3},\frac{2}{3},\frac{7}{9}\right\} &4_{\frac{14}{3},19.23}^{9,614}& 25.  \text{(prime)}\cr \hline 
2&3 & \{(2,1), ( 5,1), (5 ,1)\} & 4 &  \left\{0,\frac{2}{5},\frac{2}{5},\frac{4}{5}\right\} & 4_{\frac{28}{5},13.09}^{5,479} & 18. \text{(factors)}\cr \hline 
2&3 & \{(2,1), ( 5,1), (5 ,3)\} & 4 &   \left\{0,\frac{1}{5},\frac{2}{5},\frac{4}{5}\right\} &  4_{\frac{12}{5},5.}^{5,426} &  39.  \text{(factors)}\cr \hline 
2&3 & \{(2,1), ( 5,1), (5 ,7)\} & 4 & \left\{0,\frac{1}{5},\frac{2}{5},\frac{3}{5}\right\}   &  4_{\frac{16}{5},5.}^{5,375} & 38.  \text{(factors)}\cr \hline 
2&3 & \{(2,1), ( 5,1), (5 ,9)\} & 4 &  \left\{0,0,\frac{2}{5},\frac{3}{5}\right\}  &   
4_{0,13.09}^{5,872} &22.  \text{(factors)} \cr \hline 
2&3 & \{(2,1), ( 5,3), (5 ,3)\} & 4 &   \left\{0,\frac{3}{5},\frac{4}{5},\frac{4}{5}\right\} &  4_{\frac{36}{5},1.909}^{5,690} & 21.  \text{(factors)}\cr \hline 
2&3 & \{(2,1), ( 5,3), (5 ,7)\} & 4 &  \left\{0,0,\frac{1}{5},\frac{4}{5}\right\}  &  4_{0,1.909}^{5,456} & 23.  \text{(factors)}\cr \hline 
2&3 & \{(2,1), ( 5,3), (5 ,9)\} & 4 &  \left\{0,\frac{2}{5},\frac{3}{5},\frac{4}{5}\right\}  & 4_{\frac{24}{5},5.}^{5,223} & 41.\text{(factors)} \cr \hline 
2&3 & \{(2,1), ( 5,7), (5 ,7)\} & 4 &  \left\{0,\frac{1}{5},\frac{1}{5},\frac{2}{5}\right\}  &  4_{\frac{4}{5},1.909}^{5,248}& 20.\text{(factors)} \cr \hline 
2&3 & \{(2,1), ( 5,7), (5 ,9)\} & 4 &   \left\{0,\frac{1}{5},\frac{3}{5},\frac{4}{5}\right\} & 4_{\frac{28}{5},5.}^{5,332} & 40. \text{(factors)}\cr \hline 
2&3 & \{(2,1), ( 5,9), (5 ,9)\} & 4 &  \left\{0,\frac{1}{5},\frac{3}{5},\frac{3}{5}\right\}  & 4_{\frac{12}{5},13.09}^{5,443} & 19.\text{(factors)}\cr \hline 
2&3 & \{(3,1), ( 3,1), (5 ,1)\} & 4 &  \left\{0,\frac{1}{4},\frac{2}{5},\frac{13}{20}\right\}  & 4_{\frac{19}{5},7.236}^{20,304} & 10. \text{(factors)} \sim 31. (\text{pseudo})\cr \hline 
2&3 & \{(3,1), ( 3,1), (5 ,3)\} & 4 &  \left\{0,\frac{11}{20},\frac{3}{4},\frac{4}{5}\right\}  & 4_{\frac{33}{5},2.763}^{20,210}  &17. \text{(factors)}\cr \hline 
2&3 & \{(3,1), ( 3,1), (5 ,7)\} & 4 & \left\{0,\frac{1}{5},\frac{3}{4},\frac{19}{20}\right\}   & 4_{\frac{37}{5},2.763}^{20,210}  & 16. \text{(factors)}\cr \hline 
2&3 & \{(3,1), ( 3,1), (5 ,9)\} & 4 &  \left\{0,\frac{1}{4},\frac{3}{5},\frac{17}{20}\right\}  &  4_{\frac{31}{5},7.236}^{20,505} & 11. \text{(factors)} \sim 33. (\text{pseudo})\cr \hline 
2&3 & \{(3,1), ( 3,1), (5 ,11)\} & 4 &  \left\{0,\frac{3}{20},\frac{2}{5},\frac{3}{4}\right\}  &  4_{\frac{9}{5},7.236}^{20,451}&12. \text{(factors)} \sim 30. (\text{pseudo})\cr \hline 
2&3 & \{(3,1), ( 3,1), (5 ,13)\} & 4 &  \left\{0,\frac{1}{20},\frac{1}{4},\frac{4}{5}\right\}  &  4_{\frac{3}{5},2.763}^{20,525}&  15. \text{(factors)}\cr \hline 
2&3 & \{(3,1), ( 3,1), (5 ,17)\} & 4 & \left\{0,\frac{1}{5},\frac{1}{4},\frac{9}{20}\right\}  & 4_{\frac{7}{5},2.763}^{20,278} &  14. \text{(factors)}\cr \hline 
2&3 & \{(3,1), ( 3,1), (5 ,19)\} & 4 &  \left\{0,\frac{7}{20},\frac{3}{5},\frac{3}{4}\right\}  & 4_{\frac{21}{5},7.236}^{20,341} &  13. \text{(factors)} \sim 32. (\text{pseudo})\cr 
 \hline \hline
2 & 4 &\{(3,1), ( 3,1), (3 ,1), (5, 18)\} & 4 &   \left\{0,\frac{1}{20},\frac{4}{5}, \frac{1}{4} \right\}  & 4_{\frac{3}{5},2.763}^{20,456} &  34. \text{(factors)} =2.6 \boxtimes 2.7  
\cr \hline 
2 & 4 &\{(3,1), ( 3,1), (3 ,1), (5, 2)\}& 4 &   \left\{ 0,\frac{9}{20},\frac{1}{5},\frac{1}{4}\right\}  & 4_{\frac{7}{5},2.763}^{20,379} &  35. \text{(factors)} =2.5\boxtimes 2.7 
\cr \hline 
2 & 4 &\{(3,1), ( 3,1), (3 ,1), (5, 8)\}& 4 &   \left\{ 0,\frac{11}{20}, \frac{4}{5},\frac{3}{4}\right\}  & 4_{\frac{33}{5},2.763}^{20,771} &  36. \text{(factors)} =2.6\boxtimes  2.8
\cr \hline 
2 & 4 &\{(3,1), ( 3,1), (3 ,1), (5, 12)\}& 4 &   \left\{ 0,\frac{19}{20}, \frac{1}{5},\frac{3}{4} \right\}  & 4_{\frac{37}{5},2.763}^{20,294} &  37. \text{(factors)} =2.5\boxtimes 2.8 
\cr 
\hline\hline 
4&   3& \{(5,1), ( 5,1), (5 ,1)\}  &  4& \left\{0,\frac{1}{8},\frac{1}{8},\frac{1}{2}\right\}    &  4_{1,4.}^{8,718}&  6. \text{(prime)} \sim 44. \text{(pseudo)}\cr \hline 
4&   3& \{(5,1), ( 5,1), (5 ,3)\}  & 4&  \left\{0,\frac{1}{2},\frac{7}{8},\frac{7}{8}\right\}    & 4_{7,4.}^{8,781}  &  9. \text{(prime)} \sim 43. \text{(pseudo)}\cr \hline 
4&  3& \{(5,1), ( 5,1), (5 ,7)\}  & 4&   \left\{0,\frac{3}{8},\frac{3}{8},\frac{1}{2}\right\} & 4_{3,4.}^{8,468} &  7. \text{(prime)} \sim 45. \text{(pseudo)} \cr \hline 
4&   3& \{(5,1), ( 5,1), (5 ,13)\}  & 4&  \left\{0,\frac{1}{2},\frac{5}{8},\frac{5}{8}\right\} &  4_{5,4.}^{8,312}&  8. \text{(prime)} \sim 42. \text{(pseudo)}\cr
\hline \hline \hline
\text{Spin} (16)& 3&\{ (15,1), (15,1), (15,1) \}& 4 &  \left\{0, 0,0,\frac{1}{2}\right\}    & 4_{0,4.}^{2,750} &  1. \text{(prime, toric code)}\cr \hline 
\text{Spin} (8)& 3& \{ (7,1), (7,1), (7,1) \}  & 4 &   \left\{0,\frac{1}{2}, \frac{1}{2},\frac{1}{2} \right\}  & 4_{4,4.}^{2,250} &  2. \text{(prime)} \cr \hline 
& &  & 4 &   \left\{ 0, 0, \frac{1}{4}, \frac{3}{4}\right\}  & 4_{0,4.}^{4,375} &  3. \text{(factors )}= 2.1 \boxtimes 2.2 \cr \hline 
\text{Spin} (12)& 3& \{ (11,1), (11,1), (11,1) \} & 4 &   \left\{0, \frac{1}{2} , \frac{1}{4},  \frac{1}{4}\right\}  & 4_{2,4.}^{4,625} &  4. \text{(factors)}= 2.1 \boxtimes 2.1 \cr \hline 
\text{Spin} (20)& 3& \{ (19,1), (19,1), (19,1) \}  
& 4 &   \left\{ 0, \frac{1}{2} , \frac{3}{4},  \frac{3}{4}\right\}  &  4_{6,4.}^{4,375}&  5. \text{(factors)}= 2.2 \boxtimes 2.2 
\cr 
\hline\hline 
\end{array}
$$
\caption{MTCs from 3-manifolds: Rank 4.  The table shows the complete set of rank 4 models. For all prime models we provide a realization. \label{tab:r4}}
\end{table}


\begin{table}
$$
\begin{array}{|c|c|c|c|c|c|c|}\hline 
N & n&  M_3 (\{(p_i, q_i)\}) & \text{Rank} 
& \text{Spins} & \text{Label} & \# \cr \hline \hline
2 & 3&  \{(2,1), (3,1),  (11,1)\}& 5 & \left\{0,\frac{1}{11},\frac{2}{11},\frac{6}{11},\frac{9}{11}\right\}  & 5_{\frac{16}{11},34.64}^{11,640} & 12. \text{(prime)} \cr \hline 
2 & 3&  \{(2,1), (3,1),  (11,3)\}& 5 &  \left\{0,\frac{2}{11},\frac{3}{11},\frac{4}{11},\frac{8}{11}\right\} & 5_{\frac{20}{11},3.323}^{11,189} & 17. \text{(prime)}\cr \hline 
2 & 3&  \{(2,1), (3,1),  (11,5)\}& 5 & \left\{0,\frac{4}{11},\frac{7}{11},\frac{9}{11},\frac{10}{11}\right\}  & 5_{\frac{56}{11},9.408}^{11,540}  &14.\text{(prime)}\cr \hline 
2 & 3&  \{(2,1), (3,1),  (11,7)\}& 5 & \left\{0,\frac{4}{11},\frac{5}{11},\frac{6}{11},\frac{8}{11}\right\}  & 5_{\frac{40}{11},4.814}^{11,181} & 15.\text{(prime)}\cr \hline 
2 & 3&  \{(2,1), (3,1), (11,9)\}& 5 &   \left\{0,\frac{1}{11},\frac{5}{11},\frac{8}{11},\frac{10}{11}\right\} &  5_{\frac{80}{11},2.806}^{11,611} & 19.\text{(prime)}\cr \hline 
2 & 3&  \{(2,1), (3,1),  (11,13)\}& 5 & \left\{0,\frac{1}{11},\frac{3}{11},\frac{6}{11},\frac{10}{11}\right\}  &  5_{\frac{8}{11},2.806}^{11,238} &20.\text{(prime)}\cr \hline 
2 & 3&  \{(2,1), (3,1),  (11,15)\}& 5 & \left\{0,\frac{3}{11},\frac{5}{11},\frac{6}{11},\frac{7}{11}\right\}  & 5_{\frac{48}{11},4.814}^{11,393}  &16.\text{(prime)}\cr \hline 
2 & 3&  \{(2,1), (3,1),  (11,17)\}& 5 & \left\{0,\frac{1}{11},\frac{2}{11},\frac{4}{11},\frac{7}{11}\right\}  & 5_{\frac{32}{11},9.408}^{11,549} & 13.\text{(prime)}\cr \hline 
2 & 3&  \{(2,1), (3,1),  (11,19)\}& 5 & \left\{0,\frac{3}{11},\frac{7}{11},\frac{8}{11},\frac{9}{11}\right\}  & 5_{\frac{68}{11},3.323}^{11,508} & 18.\text{(prime)}\cr \hline 
2 & 3&  \{(2,1), (3,1),  (11,21)\}& 5 & \left\{0,\frac{2}{11},\frac{5}{11},\frac{9}{11},\frac{10}{11}\right\}  & 5_{\frac{72}{11},34.64}^{11,216} & 11. \text{(prime)}\cr \hline \hline
2 & 3&  \{(3,1), (3,1),  (6,1 )\}& 5 &  \left\{0,0,\frac{1}{8},\frac{1}{3},\frac{5}{8}\right\} &  5_{2,12.}^{24,940}  & 3. \text{(prime)} \cr \hline 
2 & 3&  \{(3,1), (3,1),  (6,5) \}& 5 & \left\{0,0,\frac{1}{8},\frac{5}{8},\frac{2}{3}\right\} & 5_{6,12.}^{24,512} & 6. \text{(prime)}\cr \hline 
2 & 3&  \{(3,1), (3,1),  (6,7 )\}& 5 & \left\{0,0,\frac{1}{3},\frac{3}{8},\frac{7}{8}\right\} & 5_{2,12.}^{24,148} & 5. \text{(prime)}\cr \hline 
2 & 3&  \{(3,1), (3,1),  (6,11) \}& 5 & \left\{0,0,\frac{3}{8},\frac{2}{3},\frac{7}{8}\right\} & 5_{6,12.}^{24,592} &  4.\text{(prime)} \cr 
\hline\hline 
2 & 3&  \{(3, 1), (3, 1),  (6, 5) \} & 5 & \left\{0,0,\frac{1}{8},\frac{5}{8},\frac{2}{3}\right\} & 5_{6,12.}^{24,273} &7.\text{(prime) same $M_3$ as  6. }
\cr     \hline
 2 & 3&  \{(3,1), (3,1 ),  (6,7) \}   & 5 &\left\{0,0,\frac{1}{3},\frac{3}{8},\frac{7}{8}\right\}  &5_{2,12.}^{24,741} & 8.\text{(prime) same $M_3$ as 5.} \cr     \hline
 2 & 3&  \{(3,1), (3,1 ),  (6,1) \}   & 5 & \left\{0,0,\frac{1}{8},\frac{1}{3},\frac{5}{8}\right\} & 5_{2,12.}^{24,615}& 9.\text{(prime) same $M_3$ as 3.} \cr     \hline
2 & 3&  \{(3,1), (3, 1),  (6,11) \}  & 5 & \left\{0,0,\frac{3}{8},\frac{2}{3},\frac{7}{8}\right\}  &5_{6,12.}^{24,814} & 10.\text{(prime) same $M_3$ as  4.} \cr     \hline
    \hline
3&3& \{(3, 1),(4, 1),(7, 1)\} & 5 & 
\{0,{ 1\over 7}, { 3\over 7}, {6 \over 7},{6 \over 7}\} & 5_{\frac{18}{7}, 35.34}^{7,101}  & 22.\text{(prime)} \cr \hline
3&3& \{(3, 1),(4, 1),(7, 2)\} & 5 & 
\{0,{ 3\over 7}, { 3\over 7}, {4 \over 7},{5 \over 7}\} &  5_{\frac{30}{7}, 4.501}^{7,125}  & 24. \text{(prime)}\cr \hline
3&3& \{(3, 1),(4, 1),(7, 4)\} & 5 & 
\{0,{ 2\over 7}, { 5\over 7}, {5 \over 7},{6 \over 7}\} &  5_{\frac{50}{7}, 2.155}^{7,255} & 26. \text{(prime)}\cr \hline
3&3& \{(3, 1),(4, 1),(7, 5)\} & 5 & 
\{0,{ 2\over 7}, { 3\over 7}, {4 \over 7},{4 \over 7}\} &  
5_{\frac{26}{7}, 4.501}^{7,408} & 23.\text{(prime)} \cr \hline
3&3& \{(3, 1),(4, 1),(7, 10)\} & 5 & 
\{0,{ 1\over 7}, { 2\over 7}, {2 \over 7},{5 \over 7}\} &  
5_{\frac{6}{7}, 2.155}^{7,342} & 25.\text{(prime)} \cr \hline
3&3& \{(3, 1),(4, 1),(7, 13)\} & 5 & 
\left\{0,{ 1\over 7}, { 1\over 7}, {4 \over 7},{6 \over 7} \right\} &  5_{\frac{38}{7}, 35.34}^{7,386} & 21.\text{(prime)} \cr 
\hline
\hline 
5& 3&  \{(6,1), (6,1), (6,1)  \}& 5 &  \left\{0,\frac{1}{5},\frac{1}{5},\frac{4}{5},\frac{4}{5}\right\} &  5_{0,5 .}^{5,110} & 1.\text{(prime)} \cr \hline 
5& 3&  \{(6,1), (6,1), (6,7)  \}& 5 &  \left\{0,\frac{2}{5},\frac{2}{5},\frac{3}{5},\frac{3}{5}\right\} & 5_{4,5}^{5,210} & 2. \text{(prime)} \cr \hline \hline
\end{array}
$$
\caption{MTCs from 3-manifolds: Rank 5. 7-10 have same spins as 3-6. and are obtained by identifying different spin 0 line as vacuum. \label{tab:Rank5}}
\end{table}

\section{Rank 6}
\label{app:r6}

In this final appendix we explore the rank 6 models. The vast majority of these are factored into rank 2 and 3 models. However we find most of the prime models, bar 10 models (out of a total of 192 rank 6 models). 
Clearly one obvious place to search for these models is to generalize the dictionary between MTCs and $M_3$, $\mathfrak{sl}(N,\C)$ data to include cases where $N$ and $q$ are not co-prime.


\begin{table}
$$
\begin{array}{|c|c|c|c|c|c|c|}\hline 
N & n&  M_3 (\{(p_i, q_i)\}) & \text{Rank} 
& \text{Spins} & \text{Label} & \# \cr \hline \hline
2 & 3&  \{(2,1),  (3, 1), ( 13, 1)\}& 6 & 
\left\{0,\frac{2}{13},\frac{4}{13},\frac{6}{13},\frac{7}{13},\frac{12}{13}\right\}& 6_{\frac{46}{13},56.74}^{13,131} & 137. (\text{prime}) \cr \hline
2 & 3& \{(2,1),  (3, 1), ( 13, 3)\}& 6 &\left\{0,\frac{2}{13},\frac{4}{13},\frac{5}{13},\frac{10}{13},\frac{11}{13}\right\} & 6_{\frac{102}{13},4.798}^{13,107} & 143. (\text{prime}) \cr \hline
2 & 3& \{(2,1),  (3, 1), ( 13, 5)\}& 6 & \left\{0,\frac{3}{13},\frac{4}{13},\frac{5}{13},\frac{6}{13},\frac{9}{13}\right\}& 6_{\frac{30}{13},3.717}^{13,162} &  145. (\text{prime}) \cr \hline
2 & 3& \{(2,1),  (3, 1), ( 13, 7)\}& 6 &\left\{0,\frac{1}{13},\frac{4}{13},\frac{8}{13},\frac{11}{13},\frac{12}{13}\right\} & 6_{\frac{14}{13},15.04}^{13,300} & 140. (\text{prime}) \cr \hline
2 & 3& \{(2,1),  (3, 1), ( 13, 9)\}& 6 &\left\{0,\frac{5}{13},\frac{6}{13},\frac{8}{13},\frac{10}{13},\frac{12}{13}\right\} & 6_{\frac{86}{13},7.390}^{13,241}  & 142. (\text{prime})  \cr \hline
2 & 3& \{(2,1),  (3, 1), ( 13, 11)\}& 6 & \left\{0,\frac{3}{13},\frac{7}{13},\frac{10}{13},\frac{11}{13},\frac{12}{13}\right\}& 6_{\frac{94}{13},3.297}^{13,764} & 147.(\text{prime})  \cr \hline
2 & 3& \{(2,1),  (3, 1), ( 13, 15)\}& 6 &\left\{0,\frac{1}{13},\frac{2}{13},\frac{3}{13},\frac{6}{13},\frac{10}{13}\right\} & 6_{\frac{10}{13},3.297}^{13,560} & 148. (\text{prime}) \cr \hline
2 & 3& \{(2,1),  (3, 1), ( 13, 17)\}& 6 & \left\{0,\frac{1}{13},\frac{3}{13},\frac{5}{13},\frac{7}{13},\frac{8}{13}\right\}&  6_{\frac{18}{13},7.390}^{13,415}&  141. (\text{prime}) \cr \hline
2 & 3& \{(2,1),  (3, 1), ( 13, 19)\}& 6 &\left\{0,\frac{1}{13},\frac{2}{13},\frac{5}{13},\frac{9}{13},\frac{12}{13}\right\} & 6_{\frac{90}{13},15.04}^{13,102} & 139. (\text{prime}) \cr \hline
2 & 3& \{(2,1),  (3, 1), ( 13, 21)\}& 6 &\left\{0,\frac{4}{13},\frac{7}{13},\frac{8}{13},\frac{9}{13},\frac{10}{13}\right\} & 6_{\frac{74}{13},3.717}^{13,481} &  146.(\text{prime}) \cr \hline
2 & 3& \{(2,1),  (3, 1), ( 13, 23)\}& 6 & \left\{0,\frac{2}{13},\frac{3}{13},\frac{8}{13},\frac{9}{13},\frac{11}{13}\right\}& 6_{\frac{2}{13},4.798}^{13,604} & 144.(\text{prime})  \cr \hline
2 & 3& \{(2,1),  (3, 1), ( 13, 25)\}& 6 &\left\{0,\frac{1}{13},\frac{6}{13},\frac{7}{13},\frac{9}{13},\frac{11}{13}\right\} & 6_{\frac{58}{13},56.74}^{13,502} &  138.(\text{prime}) \cr \hline\hline
2 & 3& \{(2,1),  (5, 1), ( 7, 1)\}& 6 &\left\{0,\frac{9}{35},\frac{2}{7},\frac{2}{5},\frac{24}{35},\frac{6}{7}\right\} & 6_{\frac{138}{35},33.63}^{35,363} & 114. (\text{factors})\cr \hline
2 & 3& \{(2,1),  (5, 1), ( 7, 3)\}& 6 & \left\{0,\frac{2}{7},\frac{2}{5},\frac{3}{7},\frac{24}{35},\frac{29}{35}\right\}& 6_{\frac{158}{35},10.35}^{35,274} & 122. (\text{factors})\cr \hline
2 & 3& \{(2,1),  (5, 1), ( 7, 5)\}& 6 & \left\{0,\frac{9}{35},\frac{2}{5},\frac{4}{7},\frac{6}{7},\frac{34}{35}\right\}& 6_{\frac{78}{35},6.661}^{35,121} & 128.(\text{factors}) \cr \hline
2 & 3& \{(2,1),  (5, 1), ( 7, 9)\}& 6 & \left\{0,\frac{1}{7},\frac{2}{5},\frac{3}{7},\frac{19}{35},\frac{29}{35}\right\}& 6_{\frac{118}{35},6.661}^{35,153} & 126.(\text{factors})\cr \hline
2 & 3& \{(2,1),  (5, 1), ( 7, 11)\}& 6 & \left\{0,\frac{4}{35},\frac{2}{5},\frac{4}{7},\frac{5}{7},\frac{34}{35}\right\}
& 6_{\frac{38}{35},10.35}^{35,423} & 124.(\text{factors}) \cr \hline
2 & 3& \{(2,1),  (5, 1), ( 7, 13)\}& 6 &\left\{0,\frac{4}{35},\frac{1}{7},\frac{2}{5},\frac{19}{35},\frac{5}{7}\right\} 
&  6_{\frac{58}{35},33.63}^{35,955}& 113.(\text{factors})\cr \hline
2 & 3& \{(2,1),  (5, 3), ( 7, 1)\}& 6 & \left\{0,\frac{3}{35},\frac{2}{7},\frac{23}{35},\frac{4}{5},\frac{6}{7}\right\}&
6_{\frac{26}{35},12.84}^{35,138} &120. (\text{factors}) \cr \hline
2 & 3& \{(2,1),  (5, 3), ( 7, 3)\}& 6 & \left\{0,\frac{3}{35},\frac{8}{35},\frac{2}{7},\frac{3}{7},\frac{4}{5}\right\} & 6_{\frac{46}{35},3.956}^{35,437} & 129. (\text{factors})\cr \hline
2 & 3& \{(2,1),  (5, 3), ( 7, 5)\}& 6 & \left\{0,\frac{13}{35},\frac{4}{7},\frac{23}{35},\frac{4}{5},\frac{6}{7}\right\}&6_{\frac{246}{35},2.544}^{35,724}  & 136. (\text{factors})\cr \hline
2 & 3& \{(2,1),  (5, 3), ( 7, 9)\}& 6 &\left\{0,\frac{1}{7},\frac{8}{35},\frac{3}{7},\frac{4}{5},\frac{33}{35}\right\} & 6_{\frac{6}{35},2.544}^{35,346} & 134.(\text{factors}) \cr \hline
2 & 3& \{(2,1),  (5, 3), ( 7, 11)\}& 6 &\left\{0,\frac{13}{35},\frac{18}{35},\frac{4}{7},\frac{5}{7},\frac{4}{5}\right\} & 6_{\frac{206}{35},3.956}^{35,723} & 131.(\text{factors}) \cr \hline
2 & 3& \{(2,1),  (5, 3), ( 7, 13)\}& 6 & \left\{0,\frac{1}{7},\frac{18}{35},\frac{5}{7},\frac{4}{5},\frac{33}{35}\right\}& 6_{\frac{226}{35},12.84}^{35,105} & 118.(\text{factors}) \cr \hline
2 & 3& \{(2,1),  (5, 7), ( 7, 1)\}& 6 &\left\{0,\frac{2}{35},\frac{1}{5},\frac{2}{7},\frac{17}{35},\frac{6}{7}\right\} & 6_{\frac{54}{35},12.84}^{35,669} & 119. (\text{factors})\cr \hline
2 & 3& \{(2,1),  (5, 7), ( 7, 3)\}& 6 &\left\{0,\frac{1}{5},\frac{2}{7},\frac{3}{7},\frac{17}{35},\frac{22}{35}\right\} &  6_{\frac{74}{35},3.956}^{35,769}& 130. (\text{factors})\cr \hline
2 & 3& \{(2,1),  (5, 7), ( 7, 5)\}& 6 & \left\{0,\frac{2}{35},\frac{1}{5},\frac{4}{7},\frac{27}{35},\frac{6}{7}\right\}& 6_{\frac{274}{35},2.544}^{35,692} &135. (\text{factors})\cr \hline
2 & 3& \{(2,1),  (5, 7), ( 7, 9)\}& 6 &\left\{0,\frac{1}{7},\frac{1}{5},\frac{12}{35},\frac{3}{7},\frac{22}{35}\right\} & 
6_{\frac{34}{35},2.544}^{35,711}&  133.(\text{factors})\cr \hline
2 & 3& \{(2,1),  (5, 7), ( 7, 11)\}& 6 &\left\{0,\frac{1}{5},\frac{4}{7},\frac{5}{7},\frac{27}{35},\frac{32}{35}\right\} &6_{\frac{234}{35},3.956}^{35,690}  & 132.(\text{factors}) \cr \hline
2 & 3& \{(2,1),  (5, 7), ( 7, 13)\}& 6 & \left\{0,\frac{1}{7},\frac{1}{5},\frac{12}{35},\frac{5}{7},\frac{32}{35}\right\}& 6_{\frac{254}{35},12.84}^{35,615} & 117. (\text{factors})\cr \hline
2 & 3& \{(2,1),  (5, 9), ( 7, 1)\}& 6 &\left\{0,\frac{2}{7},\frac{16}{35},\frac{3}{5},\frac{6}{7},\frac{31}{35}\right\} &  6_{\frac{222}{35},33.63}^{35,224}&  116.(\text{factors})\cr \hline
2 & 3& \{(2,1),  (5, 9), ( 7, 3)\}& 6 &\left\{0,\frac{1}{35},\frac{2}{7},\frac{3}{7},\frac{3}{5},\frac{31}{35}\right\} & 6_{\frac{242}{35},10.35}^{35,120} & 121.(\text{factors})\cr \hline
2 & 3& \{(2,1),  (5, 9), ( 7, 5)\}& 6 &\left\{0,\frac{6}{35},\frac{16}{35},\frac{4}{7},\frac{3}{5},\frac{6}{7}\right\} & 6_{\frac{162}{35},6.661}^{35,532} & 127. (\text{factors})\cr \hline
2 & 3& \{(2,1),  (5, 9), ( 7, 9)\}& 6 &\left\{0,\frac{1}{35},\frac{1}{7},\frac{3}{7},\frac{3}{5},\frac{26}{35}\right\} & 6_{\frac{202}{35},6.661}^{35,488} & 125. (\text{factors})\cr \hline
2 & 3& \{(2,1),  (5, 9), ( 7, 11)\}& 6 & \left\{0,\frac{6}{35},\frac{11}{35},\frac{4}{7},\frac{3}{5},\frac{5}{7}\right\}& 6_{\frac{122}{35},10.35}^{35,246} & 123. (\text{factors})\cr \hline
2 & 3& \{(2,1),  (5, 9), ( 7, 13)\}& 6 & \left\{0,\frac{1}{7},\frac{11}{35},\frac{3}{5},\frac{5}{7},\frac{26}{35}\right\}& 6_{\frac{142}{35},33.63}^{35,429} &  115. (\text{factors})\cr \hline\hline
2 & 3& \{(3,1),  (3, 1), ( 7, 1)\}& 6 &
\left\{0,\frac{1}{28},\frac{2}{7},\frac{17}{28},\frac{3}{4},\frac{6}{7}\right\}& 6_{\frac{1}{7},18.59}^{28,212} &  96. *(\text{factors})
\cr \hline
2 & 3& \{(3,1),  (3, 1), ( 7, 3)\}& 6 & \left\{0,\frac{1}{4},\frac{2}{7},\frac{3}{7},\frac{15}{28},\frac{19}{28}\right\}& 6_{\frac{19}{7},5.725}^{28,193} & 98.* (\text{factors}) 
\cr \hline
2 & 3& \{(3,1),  (3, 1), ( 7, 5)\}& 6 & \left\{0,\frac{9}{28},\frac{4}{7},\frac{17}{28},\frac{3}{4},\frac{6}{7}\right\}& 6_{\frac{45}{7},3.682}^{28,452} & 103. *(\text{factors})
\cr \hline
2 & 3& \{(3,1),  (3, 1), ( 7, 9)\}& 6 & \left\{0,\frac{1}{7},\frac{5}{28},\frac{3}{7},\frac{3}{4},\frac{25}{28}\right\}& 6_{\frac{53}{7},3.682}^{28,318}  & 101. *(\text{factors})\cr \hline
2 & 3& \{(3,1),  (3, 1), ( 7, 11)\}& 6 &\left\{0,\frac{1}{4},\frac{4}{7},\frac{5}{7},\frac{23}{28},\frac{27}{28}\right\} & 6_{\frac{51}{7},5.725}^{28,133} & 100. *  (\text{factors}) \cr \hline
2 & 3& \{(3,1),  (3, 1), ( 7, 13)\}& 6 & \left\{0,\frac{1}{7},\frac{13}{28},\frac{5}{7},\frac{3}{4},\frac{25}{28}\right\}&  6_{\frac{41}{7},18.59}^{28,114} &  95. * (\text{factors})\cr \hline

\end{array}
$$
\caption{MTCs from 3-manifolds: Rank 6. 
}
\end{table}

\begin{table}
$$
\begin{array}{|c|c|c|c|c|c|c|}\hline 
N & n&  M_3 (\{(p_i, q_i)\}) & \text{Rank} 
& \text{Spins} & \text{Label} & \# \cr \hline \hline
2 & 3& \{(3,1),  (3, 1), ( 7, 15)\}& 6 & \left\{0,\frac{3}{28},\frac{1}{4},\frac{2}{7},\frac{15}{28},\frac{6}{7}\right\}&  6_{\frac{15}{7},18.59}^{28,289} &  94. * (\text{factors})\cr \hline
2 & 3& \{(3,1),  (3, 1), ( 7, 17)\}& 6 &\left\{0,\frac{1}{28},\frac{5}{28},\frac{2}{7},\frac{3}{7},\frac{3}{4}\right\} &   6_{\frac{5}{7},5.725}^{28,424} & 97.* (\text{factors}) \cr \hline
2 & 3& \{(3,1),  (3, 1), ( 7, 19)\}& 6 &\left\{0,\frac{3}{28},\frac{1}{4},\frac{4}{7},\frac{23}{28},\frac{6}{7}\right\} &  6_{\frac{3}{7},3.682}^{28,411} &  104.* (\text{factors})\cr \hline
2 & 3& \{(3,1),  (3, 1), ( 7, 23)\}& 6 &  \left\{0,\frac{1}{7},\frac{1}{4},\frac{11}{28},\frac{3}{7},\frac{19}{28}\right\}&  6_{\frac{11}{7},3.682}^{28,112} &  102. * (\text{factors})\cr \hline
2 & 3& \{(3,1),  (3, 1), ( 7, 25)\}& 6 & \left\{0,\frac{9}{28},\frac{13}{28},\frac{4}{7},\frac{5}{7},\frac{3}{4}\right\} & 6_{\frac{37}{7},5.725}^{28,257}  &99.*  (\text{factors})\cr \hline
2 & 3& \{(3,1),  (3, 1), ( 7, 27)\}& 6 & \left\{0,\frac{1}{7},\frac{1}{4},\frac{11}{28},\frac{5}{7},\frac{27}{28}\right\} &  6_{\frac{55}{7},18.59}^{28,108} &  93.* (\text{factors})\cr \hline\hline
2 & 3 &  \{(3,1),  (4, 1), ( 5, 1)\}   & 6 & \left\{0,\frac{3}{16},\frac{2}{5},\frac{1}{2},\frac{47}{80},\frac{9}{10}\right\}    &
6_{\frac{43}{10},14.47}^{80,424}&  61.* (\text{factors}) \cr \hline 
2 & 3 &  \{(3,1),  (4, 1), ( 5, 3)\}   & 6 &  \left\{0,\frac{3}{10},\frac{39}{80},\frac{1}{2},\frac{11}{16},\frac{4}{5}\right\} & 6_{\frac{51}{10},5.527}^{80,664}&  86.* (\text{factors}) \cr \hline 
2 & 3 &  \{(3,1),  (4, 1), ( 5, 7)\}   & 6 & \left\{0,\frac{1}{5},\frac{1}{2},\frac{11}{16},\frac{7}{10},\frac{71}{80}\right\}  &6_{\frac{59}{10},5.527}^{80,423} & 85.* (\text{factors})  \cr \hline 
2 & 3 &  \{(3,1),  (4, 1), ( 5, 9)\}   & 6 &  \left\{0,\frac{1}{10},\frac{3}{16},\frac{1}{2},\frac{3}{5},\frac{63}{80}\right\} &6_{\frac{67}{10},14.47}^{80,828} &  62.* (\text{factors}) \cr \hline 
2 & 3 &  \{(3,1),  (4, 1), ( 5, 11)\}   & 6 & \left\{0,\frac{7}{80},\frac{2}{5},\frac{1}{2},\frac{11}{16},\frac{9}{10}\right\}  & 6_{\frac{3}{10},14.47}^{80,270}& 65. * (\text{factors}) \cr \hline 
2 & 3 &  \{(3,1),  (4, 1), ( 5, 13)\}   & 6 &  \left\{0,\frac{3}{16},\frac{3}{10},\frac{1}{2},\frac{4}{5},\frac{79}{80}\right\} &6_{\frac{11}{10},5.527}^{80,132} &82. * (\text{factors})  \cr \hline 
2 & 3 &  \{(3,1),  (4, 1), ( 5, 17)\}   & 6 & \left\{0,\frac{3}{16},\frac{1}{5},\frac{31}{80},\frac{1}{2},\frac{7}{10}\right\}  & 6_{\frac{19}{10},5.527}^{80,485}& 81. * (\text{factors})  \cr \hline 
2 & 3 &  \{(3,1),  (4, 1), ( 5, 19)\}   & 6 &  \left\{0,\frac{1}{10},\frac{23}{80},\frac{1}{2},\frac{3}{5},\frac{11}{16}\right\} & 6_{\frac{27}{10},14.47}^{80,528}&  66. *  (\text{factors})\cr \hline 
2 & 3 &  \{(3,1),  (4, 3), ( 5, 1)\}   & 6 &  \left\{0,\frac{1}{16},\frac{2}{5},\frac{37}{80},\frac{1}{2},\frac{9}{10}\right\} &6_{\frac{33}{10},14.47}^{80,941}&  69. * (\text{factors}) \cr \hline 
2 & 3 &  \{(3,1),  (4, 3), ( 5, 3)\}   & 6 & \left\{0,\frac{3}{10},\frac{29}{80},\frac{1}{2},\frac{9}{16},\frac{4}{5}\right\}  &6_{\frac{41}{10},5.527}^{80,589} & 80.* (\text{factors}) \cr \hline 
2 & 3 &  \{(3,1),  (4, 3), ( 5, 7)\}   & 6 & \left\{0,\frac{1}{5},\frac{1}{2},\frac{9}{16},\frac{7}{10},\frac{61}{80}\right\}& 6_{\frac{49}{10},5.527}^{80,727}& 91.*  (\text{factors})\cr \hline 
2 & 3 &  \{(3,1),  (4, 3), ( 5, 9)\}   & 6 & \left\{0,\frac{1}{16},\frac{1}{10},\frac{1}{2},\frac{3}{5},\frac{53}{80}\right\}  &6_{\frac{57}{10},14.47}^{80,544} & 73.*  (\text{factors})\cr \hline 
2 & 3 &  \{(3,1),  (4, 3), ( 5, 11)\}   & 6 & \left\{0,\frac{2}{5},\frac{1}{2},\frac{9}{16},\frac{9}{10},\frac{77}{80}\right\}  & 6_{\frac{73}{10},14.47}^{80,113}& 71.*  (\text{factors})\cr \hline 
2 & 3 &  \{(3,1),  (4, 3), ( 5, 13)\}   & 6 &  \left\{0,\frac{1}{16},\frac{3}{10},\frac{1}{2},\frac{4}{5},\frac{69}{80}\right\} & 6_{\frac{1}{10},5.527}^{80,387}& 92.* (\text{factors})\cr \hline 
2 & 3 &  \{(3,1),  (4, 3), ( 5, 17)\}   & 6 & \left\{0,\frac{1}{16},\frac{1}{5},\frac{21}{80},\frac{1}{2},\frac{7}{10}\right\}  & 6_{\frac{9}{10},5.527}^{80,787}& 79.* (\text{factors}) \cr \hline 
2 & 3 &  \{(3,1),  (4, 3), ( 5, 19)\}   & 6 &  \left\{0,\frac{1}{10},\frac{13}{80},\frac{1}{2},\frac{9}{16},\frac{3}{5}\right\} & 6_{\frac{17}{10},14.47}^{80,426}& 95.*  (\text{factors})\cr \hline \hline
2 & 3 &  \{(3,1),  (4, 5), ( 5, 1)\}   & 6 &  \left\{0,\frac{2}{5},\frac{7}{16},\frac{1}{2},\frac{67}{80},\frac{9}{10}\right\} &6_{\frac{63}{10},14.47}^{80,439} & 70.*  (\text{factors})\cr \hline 
2 & 3 &  \{(3,1),  (4, 5), ( 5, 3)\}   & 6 & \left\{0,\frac{3}{10},\frac{1}{2},\frac{59}{80},\frac{4}{5},\frac{15}{16}\right\}  & 6_{\frac{71}{10},5.527}^{80,102}&  90.* (\text{factors})\cr \hline 
2 & 3 &  \{(3,1),  (4, 5), ( 5, 7)\}   & 6 &  \left\{0,\frac{11}{80},\frac{1}{5},\frac{1}{2},\frac{7}{10},\frac{15}{16}\right\} & 6_{\frac{79}{10},5.527}^{80,135}& 77. * (\text{factors}) \cr \hline 
2 & 3 &  \{(3,1),  (4, 5), ( 5, 9)\}   & 6 &  \left\{0,\frac{3}{80},\frac{1}{10},\frac{7}{16},\frac{1}{2},\frac{3}{5}\right\} & 6_{\frac{7}{10},14.47}^{80,191}& 74.*  (\text{factors})\cr \hline 
2 & 3 &  \{(3,1),  (4, 5), ( 5, 11)\}   & 6 & \left\{0,\frac{27}{80},\frac{2}{5},\frac{1}{2},\frac{9}{10},\frac{15}{16}\right\}  & 6_{\frac{23}{10},14.47}^{80,675}& 72.* (\text{factors}) \cr \hline 
2 & 3 &  \{(3,1),  (4, 5), ( 5, 13)\}   & 6 &  \left\{0,\frac{19}{80},\frac{3}{10},\frac{7}{16},\frac{1}{2},\frac{4}{5}\right\} & 6_{\frac{31}{10},5.527}^{80,478}& 78. * (\text{factors}) \cr \hline 
2 & 3 &  \{(3,1),  (4, 5), ( 5, 17)\}   & 6 & \left\{0,\frac{1}{5},\frac{7}{16},\frac{1}{2},\frac{51}{80},\frac{7}{10}\right\}  &6_{\frac{39}{10},5.527}^{80,489} & 89. * (\text{factors})\cr \hline 
2 & 3 &  \{(3,1),  (4, 5), ( 5, 19)\}   & 6 &  \left\{0,\frac{1}{10},\frac{1}{2},\frac{43}{80},\frac{3}{5},\frac{15}{16}\right\} & 6_{\frac{47}{10},14.47}^{80,143}& 76.*  (\text{factors})\cr \hline \hline
2 & 3 &  \{(3,1),  (4, 7), ( 5, 1)\}   & 6 & \left\{0,\frac{5}{16},\frac{2}{5},\frac{1}{2},\frac{57}{80},\frac{9}{10}\right\}  & 6_{\frac{53}{10},14.47}^{80,884}& 63.*  (\text{factors})\cr \hline 
2 & 3 &  \{(3,1),  (4, 7), ( 5, 3)\}   & 6 & \left\{0,\frac{3}{10},\frac{1}{2},\frac{49}{80},\frac{4}{5},\frac{13}{16}\right\}  & 6_{\frac{61}{10},5.527}^{80,785}& 88. * (\text{factors}) \cr \hline 
2 & 3 &  \{(3,1),  (4, 7), ( 5, 7)\}   & 6 & \left\{0,\frac{1}{80},\frac{1}{5},\frac{1}{2},\frac{7}{10},\frac{13}{16}\right\}  &6_{\frac{69}{10},5.527}^{80,529} & 87. * (\text{factors}) \cr \hline 
2 & 3 &  \{(3,1),  (4, 7), ( 5, 9)\}   & 6 &  \left\{0,\frac{1}{10},\frac{5}{16},\frac{1}{2},\frac{3}{5},\frac{73}{80}\right\} &6_{\frac{77}{10},14.47}^{80,657} & 64.*  (\text{factors})\cr \hline 
2 & 3 &  \{(3,1),  (4, 7), ( 5, 11)\}   & 6 &  \left\{0,\frac{17}{80},\frac{2}{5},\frac{1}{2},\frac{13}{16},\frac{9}{10}\right\} &6_{\frac{13}{10},14.47}^{80,621} & 67.*  (\text{factors})\cr \hline 
2 & 3 &  \{(3,1),  (4, 7), ( 5, 13)\}   & 6 &  \left\{0,\frac{9}{80},\frac{3}{10},\frac{5}{16},\frac{1}{2},\frac{4}{5}\right\} & 6_{\frac{21}{10},5.527}^{80,400}&  84.* (\text{factors})\cr \hline 
2 & 3 &  \{(3,1),  (4, 7), ( 5, 17)\}   & 6 & \left\{0,\frac{1}{5},\frac{5}{16},\frac{1}{2},\frac{41}{80},\frac{7}{10}\right\}  & 6_{\frac{29}{10},5.527}^{80,495}& 83.* (\text{factors})\cr \hline 
2 & 3 &  \{(3,1),  (4, 7), ( 5, 19)\}   & 6 &  \left\{0,\frac{1}{10},\frac{33}{80},\frac{1}{2},\frac{3}{5},\frac{13}{16}\right\} & 6_{\frac{37}{10},14.47}^{80,629}& 68.* (\text{factors}) \cr \hline 
\hline 
\end{array}
$$
\caption{MTCs from 3-manifolds: Rank 6. (cont.) }
\end{table}

\begin{table}
$$
\begin{array}{|c|c|c|c|c|c|c|}\hline 
N & n&  M_3 (\{(p_i, q_i)\}) & \text{Rank} 
& \text{Spins} & \text{Label} & \# \cr \hline \hline
3&3& \{(4, 1),(4, 1),(5, 1)\} & 6 & 
\{0,\frac{1}{3},\frac{1}{3},\frac{3}{5},\frac{14}{15},\frac{14}{15}\} &  
6_{\frac{36}{5}, 10.85}^{15,166}
& 22. (\text{factors}) \cr \hline
3&3& \{(4, 1),(4, 1),(5, 4)\} & 6 & 
\{0,\frac{1}{3},\frac{1}{3},\frac{2}{5},\frac{11}{15},\frac{11}{15}\} &  
6_{\frac{24}{5}, 10.85}^{15,257}
& 21. (\text{factors})\cr \hline
3&3& \{(4, 1),(4, 1),(5, 7)\} & 6 & 
\{0,\frac{2}{15},\frac{2}{15},\frac{1}{3},\frac{1}{3},\frac{4}{5}\} &  
6_{\frac{8}{5}, 4.145}^{15,481}
& 26. (\text{factors})\cr \hline
3&3& \{(4, 1),(4, 1),(5, 13)\} & 6 & 
\{0,\frac{1}{5},\frac{1}{3},\frac{1}{3},\frac{8}{15},\frac{8}{15}\} &  
6_{\frac{12}{5}, 4.145}^{15,440}
& 25. (\text{factors})\cr \hline
3&3& \{(4, 1),(4, 5),(5, 1)\} & 6 & 
\{ 0,\frac{4}{15},\frac{4}{15},\frac{3}{5},\frac{2}{3},\frac{2}{3}\} &  
6_{\frac{16}{5}, 10.85}^{15,262} &  24. (\text{factors})\cr \hline
3&3& \{(4, 1),(4, 5),(5, 4)\} & 6 & 
\{ 0,\frac{1}{15},\frac{1}{15},\frac{2}{5},\frac{2}{3},\frac{2}{3}\} &  
6_{\frac{4}{5}, 10.85}^{15,801}& 23.(\text{factors}) \cr \hline
3&3& \{(4, 1),(4, 5),(5,7 )\} & 6 & 
\{ 0,\frac{7}{15},\frac{7}{15},\frac{2}{3},\frac{2}{3},\frac{4}{5}\} &  
6_{\frac{28}{5}, 4.145}^{15,623} & 28. (\text{factors})\cr \hline
3&3& \{(4,1 ),(4, 5),(5,13  )\} & 6 & 
\{ 0,\frac{1}{5},\frac{2}{3},\frac{2}{3},\frac{13}{15},\frac{13}{15}\} &  
6_{\frac{32}{5}, 4.145}^{15,350}&  27.(\text{factors}) \cr \hline
\hline 
2& 4& \{(3,1), (3,1), (4,3), (5,6)\}& 6& \left\{0,\frac{1}{16},\frac{2}{5},\frac{37}{80},\frac{1}{2},\frac{9}{10}\right\}& 6_{\frac{33}{10},14.47}^{80,798}& 29.  (\text{factors}) \cr \hline 
 2& 4& \{(3,1), (3,1), (4,3), (5,14)\}& 6& \left\{0,\frac{1}{16},\frac{1}{10},\frac{1}{2},\frac{3}{5},\frac{53}{80}\right\} & 6_{\frac{57}{10},14.47}^{80,376} &  30. (\text{factors}) \cr \hline 
 2& 4& \{(3,1), (3,1), (4,5), (5,6)\}& 6& \left\{0,\frac{2}{5},\frac{7}{16},\frac{1}{2},\frac{67}{80},\frac{9}{10}\right\} & 6_{\frac{63}{10},14.47}^{80,146} & 31.    (\text{factors}) \cr \hline 
 2& 4& \{(3,1), (3,1), (4,5), (5,14)\}& 6& \left\{0,\frac{3}{80},\frac{1}{10},\frac{7}{16},\frac{1}{2},\frac{3}{5}\right\} & 6_{\frac{7}{10},14.47}^{80,111} &32.    (\text{factors}) \cr \hline 
 2& 4& \{(3,1), (3,1), (4,3 ), (5, 16)\}& 6& \left\{0,\frac{2}{5},\frac{1}{2},\frac{9}{16},\frac{9}{10},\frac{77}{80}\right\} &  6_{\frac{73}{10},14.47}^{80,215}&  33.  (\text{factors}) \cr \hline 
 2& 4& \{(3,1), (3,1), (4,3 ), (5,4 )\}& 6& \left\{0,\frac{1}{10},\frac{13}{80},\frac{1}{2},\frac{9}{16},\frac{3}{5}\right\} & 6_{\frac{17}{10},14.47}^{80,878 }& 34.  (\text{factors}) \cr \hline 

 2& 4& \{(3,1), (3,1), (4,5 ), (5,16 )\}& 6& \left\{0,\frac{27}{80},\frac{2}{5},\frac{1}{2},\frac{9}{10},\frac{15}{16}\right\} & 6_{\frac{23}{10},14.47}^{80,108} &  35.  (\text{factors}) \cr \hline 
 2& 4& \{(3,1), (3,1), (4,5 ), (5,4 )\}& 6& \left\{0,\frac{1}{10},\frac{1}{2},\frac{43}{80},\frac{3}{5},\frac{15}{16}\right\} &  6_{\frac{47}{10},14.47}^{80,518}&36.   (\text{factors}) \cr \hline 
 2& 4& \{(3,1), (3,1), (4, 1), (5, 6)\}& 6& \left\{0,\frac{3}{16},\frac{2}{5},\frac{1}{2},\frac{47}{80},\frac{9}{10}\right\} &  6_{\frac{43}{10},14.47}^{80,596}&  37. (\text{factors}) \cr \hline 
2& 4& \{(3,1), (3,1), (4, 7), (5,6 )\}& 6& \left\{0,\frac{5}{16},\frac{2}{5},\frac{1}{2},\frac{57}{80},\frac{9}{10}\right\} & 6_{\frac{53}{10},14.47}^{80,136} &  38. (\text{factors}) \cr \hline 
 2& 4& \{(3,1), (3,1), (4, 1), (5, 16)\}& 6& \left\{0,\frac{7}{80},\frac{2}{5},\frac{1}{2},\frac{11}{16},\frac{9}{10}\right\} &6_{\frac{3}{10},14.47}^{80,750}  & 39.  (\text{factors}) \cr \hline 
  2& 4& \{(3,1), (3,1), (4, 7), (5, 16)\}& 6& \left\{0,\frac{17}{80},\frac{2}{5},\frac{1}{2},\frac{13}{16},\frac{9}{10}\right\} & 6_{\frac{13}{10},14.47}^{80,399} &  40. (\text{factors}) \cr \hline 
 2& 4& \{(3,1), (3,1), (4,1 ), (5,14 )\}& 6& \left\{0,\frac{1}{10},\frac{3}{16},\frac{1}{2},\frac{3}{5},\frac{63}{80}\right\} &  6_{\frac{67}{10},14.47}^{80,100}& 41.  (\text{factors}) \cr \hline 
  2& 4& \{(3,1), (3,1), (4,7 ), (5, 14)\}& 6&  \left\{0,\frac{1}{10},\frac{5}{16},\frac{1}{2},\frac{3}{5},\frac{73}{80}\right\}& 6_{\frac{77}{10},14.47}^{80,157} &   42.(\text{factors}) \cr \hline 
  2& 4& \{(3,1), (3,1), (4, 1), (5,4 )\}& 6& \left\{0,\frac{1}{10},\frac{23}{80},\frac{1}{2},\frac{3}{5},\frac{11}{16}\right\} & 6_{\frac{27}{10},14.47}^{80,392} &   43.(\text{factors}) \cr \hline 
 2& 4& \{(3,1), (3,1), (4,7 ), (5, 4)\}& 6& \left\{0,\frac{1}{10},\frac{33}{80},\frac{1}{2},\frac{3}{5},\frac{13}{16}\right\} &  6_{\frac{37}{10},14.47}^{80,857}&  44. (\text{factors}) \cr \hline 
  2& 4& \{(3,1), (3,1), (4,7 ), (5, 12)\}& 6& \left\{0,\frac{1}{80},\frac{1}{5},\frac{1}{2},\frac{7}{10},\frac{13}{16}\right\} & 6_{\frac{69}{10},5.527}^{80,106} &  45. (\text{factors}) \cr \hline   
  2& 4& \{(3,1), (3,1), (4, 7), (5,18 )\}& 6& \left\{0,\frac{9}{80},\frac{3}{10},\frac{5}{16},\frac{1}{2},\frac{4}{5}\right\} & 6_{\frac{21}{10},5.527}^{80,384} &  46. (\text{factors}) \cr \hline 
 2& 4& \{(3,1), (3,1), (4, 1), (5, 2)\}& 6& \left\{0,\frac{3}{16},\frac{1}{5},\frac{31}{80},\frac{1}{2},\frac{7}{10}\right\} &  6_{\frac{19}{10},5.527}^{80,519}&  47. (\text{factors}) \cr \hline 
  2& 4& \{(3,1), (3,1), (4, 1), (5,8 )\}& 6& \left\{0,\frac{3}{10},\frac{39}{80},\frac{1}{2},\frac{11}{16},\frac{4}{5}\right\} & 6_{\frac{51}{10},5.527}^{80,717} &  48. (\text{factors}) \cr \hline   
  2& 4& \{(3,1), (3,1), (4, 3), (5, 2)\}& 6& \left\{0,\frac{1}{16},\frac{1}{5},\frac{21}{80},\frac{1}{2},\frac{7}{10}\right\} & 6_{\frac{9}{10},5.527}^{80,458} &  49. (\text{factors}) \cr \hline 
 2& 4& \{(3,1), (3,1), (4, 3), (5, 18)\}& 6& \left\{0,\frac{1}{16},\frac{3}{10},\frac{1}{2},\frac{4}{5},\frac{69}{80}\right\} & 6_{\frac{1}{10},5.527}^{80,117} &  50.  (\text{factors}) \cr \hline 
  2& 4& \{(3,1), (3,1), (4, 5), (5, 2)\}& 6& \left\{0,\frac{1}{5},\frac{7}{16},\frac{1}{2},\frac{51}{80},\frac{7}{10}\right\} & 6_{\frac{39}{10},5.527}^{80,488} &  51.  (\text{factors}) \cr \hline  
    2& 4& \{(3,1), (3,1), (4, 5), (5, 18)\}& 6& \left\{0,\frac{19}{80},\frac{3}{10},\frac{7}{16},\frac{1}{2},\frac{4}{5}\right\} & 6_{\frac{31}{10},5.527}^{80,305} &  52.  (\text{factors}) \cr \hline 
 2& 4& \{(3,1), (3,1), (4, 3), (5,12 )\}& 6&    \left\{0,\frac{1}{5},\frac{1}{2},\frac{9}{16},\frac{7}{10},\frac{61}{80}\right\}& 6_{\frac{49}{10},5.527}^{80,464}& 53. (\text{factors}) \cr \hline 
  2& 4& \{(3,1), (3,1), (4, 3), (5, 8)\}& 6& \left\{0,\frac{3}{10},\frac{29}{80},\frac{1}{2},\frac{9}{16},\frac{4}{5}\right\} & 6_{\frac{41}{10},5.527}^{80,194} &  54. (\text{factors}) \cr \hline   
  2& 4& \{(3,1), (3,1), (4, 5), (5, 12 )\}& 6&   \left\{0,\frac{11}{80},\frac{1}{5},\frac{1}{2},\frac{7}{10},\frac{15}{16}\right\} &6_{\frac{79}{10},5.527}^{80,822} &  55. (\text{factors}) \cr \hline 
 2& 4& \{(3,1), (3,1), (4, 5), (5, 8)\}& 6& \left\{0,\frac{3}{10},\frac{1}{2},\frac{59}{80},\frac{4}{5},\frac{15}{16}\right\} & 6_{\frac{71}{10},5.527}^{80,240} &  56.  (\text{factors}) \cr \hline 
  2& 4& \{(3,1), (3,1), (4, 7), (5, 2)\}& 6& \left\{0,\frac{1}{5},\frac{5}{16},\frac{1}{2},\frac{41}{80},\frac{7}{10}\right\} & 6_{\frac{29}{10},5.527}^{80,420}&  57. (\text{factors}) \cr \hline   
  2& 4& \{(3,1), (3,1), (4, 7), (5,8 )\}& 6& \left\{0,\frac{3}{10},\frac{1}{2},\frac{49}{80},\frac{4}{5},\frac{13}{16}\right\} & 6_{\frac{61}{10},5.527}^{80,862} &  58. (\text{factors}) \cr \hline 
 2& 4& \{(3,1), (3,1), (4, 1), (5,12 )\}& 6& \left\{0,\frac{1}{5},\frac{1}{2},\frac{11}{16},\frac{7}{10},\frac{71}{80}\right\} & 6_{\frac{59}{10},5.527}^{80,113} &   59.(\text{factors}) \cr \hline 
  2& 4& \{(3,1), (3,1), (4, 1), (5, 18)\}& 6& \left\{0,\frac{3}{16},\frac{3}{10},\frac{1}{2},\frac{4}{5},\frac{79}{80}\right\} & 6_{\frac{11}{10},5.527}^{80,545}&  60. (\text{factors}) \cr   
  \hline\hline 
\end{array}
$$
\caption{MTCs from 3-manifolds: Rank 6. (cont.) }
\end{table}

\begin{table}
$$
\begin{array}{|c|c|c|c|c|c|c|}\hline 
N & n&  M_3 (\{(p_i, q_i)\}) & \text{Rank} 
& \text{Spins} & \text{Label} & \# \cr \hline \hline
2& 4& \{(3,1), (3,1), (3, 1), (7, 6)\}& 6&  \left\{0,\frac{1}{7},\frac{1}{4},\frac{11}{28},\frac{5}{7},\frac{27}{28}\right\}& 6_{\frac{55}{7},18.59}^{28,124}&  159. (\text{pseudo}) \cr \hline   
2& 4& \{(3,1), (3,1), (3, 1), (7, 16)\}& 6& \left\{0,\frac{1}{7},\frac{5}{28},\frac{3}{7},\frac{3}{4},\frac{25}{28}\right\} &6_{\frac{53}{7},3.682}^{28,127}&  168. (\text{factors}) \cr \hline   
2& 4& \{(3,1), (3,1), (3, 1), (7, 8)\}& 6& \left\{0,\frac{1}{28},\frac{2}{7},\frac{17}{28},\frac{3}{4},\frac{6}{7}\right\} & 6_{\frac{1}{7},18.59}^{28,277}&  162. (\text{factors}) \cr \hline   
2& 4& \{(3,1), (3,1), (3, 1), (7, 4)\}& 6&  \left\{0,\frac{9}{28},\frac{13}{28},\frac{4}{7},\frac{5}{7},\frac{3}{4}\right\}&6_{\frac{37}{7},5.725}^{28,806} & 165.  (\text{factors}) \cr \hline   
2& 4& \{(3,1), (3,1), (3, 1), (7, 2)\}& 6& \left\{0,\frac{1}{7},\frac{1}{4},\frac{11}{28},\frac{3}{7},\frac{19}{28}\right\} & 6_{\frac{11}{7},3.682}^{28,782} & 167.  (\text{factors}) \cr \hline   
2& 4& \{(3,1), (3,1), (3, 1), (7, 10)\}& 6& \left\{0,\frac{1}{4},\frac{2}{7},\frac{3}{7},\frac{15}{28},\frac{19}{28}\right\} &6_{\frac{19}{7},5.725}^{28,399} &   164. (\text{factors}) \cr \hline   
2& 4& \{(3,1), (3,1), (3, 1), (7, 12 )\}& 6& \left\{0,\frac{9}{28},\frac{4}{7},\frac{17}{28},\frac{3}{4},\frac{6}{7}\right\} & 6_{\frac{45}{7},3.682}^{28,756}&   170. (\text{factors}) \cr \hline   
2& 4& \{(3,1), (3,1), (3, 1), (7, 18)\}& 6& \left\{0,\frac{1}{4},\frac{4}{7},\frac{5}{7},\frac{23}{28},\frac{27}{28}\right\} & 6_{\frac{51}{7},5.725}^{28,267}&  166.   (\text{factors}) \cr \hline   
2& 4& \{(3,1), (3,1), (3, 1), (7, 20)\}& 6& \left\{0,\frac{1}{7},\frac{13}{28},\frac{5}{7},\frac{3}{4},\frac{25}{28}\right\} &6_{\frac{41}{7},18.59}^{28,130} &   160. (\text{factors})\cr \hline   
2& 4& \{(3,1), (3,1), (3, 1), (7, 11)\}& 6& \left\{0,\frac{3}{28},\frac{1}{4},\frac{2}{7},\frac{15}{28},\frac{6}{7}\right\} &6_{\frac{15}{7},18.59}^{28,779} &  161. (\text{factors})\cr \hline   
2& 4& \{(3,1), (3,1), (3, 1), (7,24 )\}& 6& \left\{0,\frac{1}{28},\frac{5}{28},\frac{2}{7},\frac{3}{7},\frac{3}{4}\right\} &6_{\frac{5}{7},5.725}^{28,578}&  163.(\text{factors})\cr \hline   
2& 4& \{(3,1), (3,1), (3, 1), (7, 26)\}& 6& \left\{0,\frac{3}{28},\frac{1}{4},\frac{4}{7},\frac{23}{28},\frac{6}{7}\right\} & 6_{\frac{3}{7},3.682}^{28,797}& 169.  (\text{factors})\cr \hline  \hline

\text{Spin} (8)& 3& \{ (6,1), (7, 1), (9,17) \} & 4 &   \left\{0,
\frac{1}{9},\frac{1}{9},\frac{1}{9}, \frac{1}{3}, \frac{2}{3} \right\}  & 6_{\frac{8}{3},74.61}^{9,186} & 149. (\text{prime}) \cr \hline 
\text{Spin} (8)& 3& \{ (6,1), (7,1), (9,1) \} & 4 &   \left\{ 0,
\frac{8}{9},\frac{8}{9},\frac{8}{9}, \frac{2}{3}, \frac{1}{3} \right\}  & 6_{\frac{16}{3},74.61}^{9,452} & 150. (\text{prime})\cr \hline 
\text{Spin} (8)& 3& \{ (6,1), (7,1), (9,11) \} & 4 &   \left\{ 0,
\frac{4}{9},\frac{4}{9},\frac{4}{9}, \frac{2}{3}, \frac{1}{3} \right\}  & 6_{\frac{8}{3},3.834}^{9,226} & 151. (\text{prime})\cr \hline 
\text{Spin} (8)& 3& \{ (6,1), (7,1), (9,7) \} & 4 &   \left\{ 0,
\frac{5}{9},\frac{5}{9},\frac{5}{9}, \frac{2}{3}, \frac{1}{3} \right\}  & 6_{\frac{16}{3},3.834}^{9,342} & 152. (\text{prime})\cr \hline 
\text{Spin} (8)& 3& \{ (6,1), (7,1), (9,13) \} & 4 &   \left\{ 0,
\frac{2}{9},\frac{2}{9},\frac{2}{9}, \frac{2}{3}, \frac{1}{3} \right\}  & 6_{\frac{4}{3},2.548}^{9,789} & 153.(\text{prime}) \cr \hline 
\text{Spin} (8)& 3& \{ (6,1), (7,1), (9,5) \} & 4 &   \left\{ 0,
\frac{7}{9},\frac{7}{9},\frac{7}{9}, \frac{2}{3}, \frac{1}{3} \right\}  & 6_{\frac{20}{3},2.548}^{9,632} & 154.(\text{prime}) \cr 
\hline  \hline
E_7 & 3  & \{ (18,1), (19,1), (21, 5) \}& 6& \left\{  0,0,\frac{1}{7},\frac{2}{7},\frac{4}{7},\frac{2}{3} \right\}& 6_{6,100.6}^{21,154} &  155. (\text{prime})  
\cr \hline 
E_7 & 3  & \{ (18,1), (19,1), (21,1) \}& 6& \left\{  0, 0, \frac{3}{7},\frac{5}{7}, \frac{6}{7},\frac{1}{3} \right\}& 6_{2,100.6}^{21,320} & 156. (\text{prime}) \cr \hline 
E_7 & 3  & \{ (18,1), (19,1), (21,19) \}& 6& \left\{ 0, 0,\frac{1}{3},\frac{1}{7},\frac{2}{7},\frac{4}{7}  \right\}& 6_{2,4.382}^{21,215} &   157.(\text{prime}) \cr \hline 
E_7 & 3  & \{ (18,1), (19,1), (21,11) \}& 6& \left\{ 0,0,\frac{2}{3},\frac{3}{7},\frac{5}{7},\frac{6}{7}  \right\}&6_{6,4.382}^{21,604} & 158. (\text{prime}) \cr \hline
\hline
\end{array}
$$
\caption{MTCs from 3-manifolds: Rank 6. (cont.) }
\end{table}

\begin{table}
$$
\begin{array}{|c|c|c|c|}\hline 
\text{Rank} 
& \text{Spins} & \text{Label} & \# \cr \hline \hline
  6& \left\{ 0,\frac{1}{9},\frac{2}{3},\frac{4}{9}, \frac{1}{3},\frac{7}{9}\right\} &6_{\frac{16}{3},9.}^{9,746} &  171. (\text{prime})\cr \hline 
  6& \left\{ 0,\frac{8}{9},\frac{1}{3},\frac{5}{9},\frac{2}{3}, \frac{2}{9} \right\} & 6_{\frac{8}{3},9.}^{9,711} &  172. (\text{prime})\cr \hline \hline 
  6& \left\{  0,0,0, \frac{1}{5},\frac{4}{5},\frac{1}{2}  \right\}&6_{0,20.}^{10,699}  &  105.(\text{prime}) \cr \hline  
  6& \left\{ 0,0,0, \frac{2}{5},\frac{3}{5},\frac{1}{2} \right\}& 6_{4,20.}^{10,990} & 106. (\text{prime}) \cr \hline  
6& \left\{ 0, 0, 0, \frac{2}{5},\frac{3}{5},\frac{1}{2}   \right\}& 6_{4,20.}^{10,101} & 107. (\text{prime}) ,\text{{same spin as 106.}} \cr \hline  
6& \left\{   0, 0,0,\frac{1}{5},\frac{4}{5},\frac{1}{2}  \right\}& 6_{0,20.}^{10,419} & 108. (\text{prime}),   \text{{same spin as 105.}}\cr \hline  
  6& \left\{ 0, 0, \frac{1}{5},\frac{4}{5},\frac{1}{4},\frac{3}{4}  \right\}&6_{0,20.}^{20,139}  &  109.(\text{prime}) \cr \hline  
 6& \left\{  0,0,\frac{2}{5},\frac{3}{5},\frac{1}{4},\frac{3}{4}  \right\}& 6_{4,20.}^{20,739} &  110. (\text{prime}) \cr \hline  
 6& \left\{ 0,0,\frac{2}{5},\frac{3}{5}, \frac{1}{4}, \frac{3}{4}  \right\}& 6_{4,20.}^{20,180} & 111. (\text{prime}),  \text{{same spin as 110.}} \cr \hline  
  6& \left\{  0,0,\frac{1}{5},\frac{4}{5},\frac{1}{4},\frac{3}{4} \right\}& 6_{0,20.}^{20,419} & 112. (\text{prime})\text{, {same spins as 109.}} \cr \hline  \hline
\end{array}
$$
\caption{MTCs from 3-manifolds: Rank 6: missing Prime models }
\end{table}

\begin{table}
$$
\begin{array}{|c|c|c|c|}\hline 
\text{Rank} 
& \text{Spins} & \text{Label} & \# \cr \hline \hline
6& \left\{  0,\frac{1}{3},\frac{1}{3},\frac{1}{4},\frac{7}{12},\frac{7}{12}  \right\}&  6_{3,6.}^{12,534}&  1. (\text{factors}) \cr \hline  
6 & \left\{ 0,\frac{1}{3},\frac{1}{3},\frac{3}{4},\frac{1}{12},\frac{1}{12}   \right\}&  6_{1,6.}^{12,701}& 2. (\text{factors}) \cr \hline  
6 & \left\{  0,\frac{2}{3},\frac{2}{3},\frac{1}{4},\frac{11}{12},\frac{11}{12}  \right\}& 6_{7,6.}^{12,113} &  3.(\text{factors}) \cr \hline  
6 & \left\{ 0,\frac{2}{3},\frac{2}{3},\frac{3}{4},\frac{5}{12},\frac{5}{12}  \right\}&  6_{5,6.}^{12,298} &  4. (\text{factors})\cr \hline  
6 & \left\{ 0,\frac{1}{2},\frac{1}{4},\frac{3}{4},\frac{1}{16},\frac{5}{16}   \right\}&  6_{\frac{3}{2},8.}^{16,688}& 5. (\text{factors}) \cr \hline  
6 & \left\{  0,\frac{1}{2},\frac{1}{4},\frac{3}{4},\frac{3}{16},\frac{7}{16}  \right\}& 6_{\frac{5}{2},8.}^{16,511} &  6.(\text{factors}) \cr \hline  
6 & \left\{ 0,\frac{1}{2}, \frac{1}{4},\frac{3}{4}, \frac{9}{16},\frac{13}{16}   \right\}& 6_{\frac{11}{2},8.}^{16,548} &  7.(\text{factors}) \cr \hline  
6 & \left\{  0,\frac{1}{2},
\frac{1}{4},\frac{3}{4},
\frac{11}{16},\frac{15}{16} \right\}& 6_{\frac{13}{2},8.}^{16,107} &  8.(\text{factors}) \cr \hline  
6 & \left\{  0,\frac{1}{2},\frac{1}{4},\frac{3}{4},\frac{1}{16},\frac{13}{16}  \right\}& 6_{\frac{15}{2},8.}^{16,357} &  9.(\text{factors}) \cr \hline  
6 & \left\{  0,\frac{1}{2},
\frac{1}{4},\frac{3}{4},\frac{3}{16},\frac{15}{16} \right\}& 6_{\frac{1}{2},8.}^{16,710} &  10.(\text{factors}) \cr \hline  
6 & \left\{ 0,\frac{1}{2},\frac{1}{4},\frac{3}{4},\frac{5}{16},\frac{9}{16}  \right\}& 6_{\frac{7}{2},8.}^{16,346} &  11.(\text{factors}) \cr \hline  
6 & \left\{  0 ,\frac{1}{2} ,\frac{1}{4},\frac{3}{4},
\frac{7}{16},\frac{11}{16}  \right\}& 6_{\frac{9}{2},8.}^{16,357} &  12.(\text{factors}) \cr \hline  
6 & \left\{ 0,\frac{1}{2},\frac{1}{4},\frac{3}{4},
\frac{1}{16},\frac{13}{16}  \right\}& 6_{\frac{15}{2},8.}^{16,107} &  13.(\text{factors}) \cr \hline  
6 & \left\{  0,\frac{1}{2},\frac{1}{4},\frac{3}{4},
\frac{3}{16},\frac{15}{16}  \right\}& 6_{\frac{1}{2},8.}^{16,460} &  14.(\text{factors}) \cr \hline  
6 & \left\{   0,\frac{1}{2},\frac{1}{4},\frac{3}{4},\frac{5}{16},\frac{9}{16} \right\}& 6_{\frac{7}{2},8.}^{16,246} &  15.(\text{factors}) \cr \hline  
6 & \left\{  0,\frac{1}{2},\frac{1}{4},\frac{3}{4},\frac{7}{16},\frac{11}{16} \right\}&  6_{\frac{9}{2},8.}^{16,107}&  16.(\text{factors}) \cr \hline  
6 & \left\{  0,\frac{1}{2},
\frac{1}{4},\frac{3}{4},
\frac{1}{16},\frac{5}{16}  \right\}&  6_{\frac{3}{2},8.}^{16,438}&  17.(\text{factors}) \cr \hline  
6 & \left\{ 0,\frac{1}{2},\frac{1}{4},\frac{3}{4},\frac{3}{16},\frac{7}{16}  \right\}&  6_{\frac{5}{2},8.}^{16,261}&  18.(\text{factors}) \cr \hline  
6 & \left\{ 0,\frac{1}{2},\frac{1}{4},\frac{3}{4},
\frac{9}{16},\frac{13}{16}   \right\}& 6_{\frac{11}{2},8.}^{16,798} &  19.(\text{factors}) \cr \hline  
6 & \left\{  0,\frac{1}{2},\frac{1}{4},\frac{3}{4},
\frac{11}{16},\frac{15}{16} \right\}& 6_{\frac{13}{2},8.}^{16,132} &  20.(\text{factors}) \cr \hline  \hline
6 & \left\{  0,\frac{1}{2},\frac{1}{16},\frac{1}{4},
\frac{3}{4},\frac{5}{16}  \right\}&  6_{\frac{3}{2},8.}^{16,507} &  173.(\text{factors}) \cr \hline 
6 & \left\{   0,\frac{1}{2},\frac{1}{16},\frac{1}{4},
\frac{3}{4},\frac{13}{16} \right\}& 6_{\frac{15}{2},8.}^{16,130} &  174.(\text{factors}) \cr \hline 
6 & \left\{  0,\frac{1}{2},\frac{7}{16},\frac{1}{4},
\frac{3}{4},\frac{3}{16}  \right\}& 6_{\frac{5}{2},8.}^{16,154} &  175.(\text{factors}) \cr \hline 
6 & \left\{0,\frac{1}{2},\frac{7}{16},\frac{1}{4},\frac{3}{4},\frac{11}{16} \right\}& 6_{\frac{9}{2},8.}^{16,772} &  176.(\text{factors}) \cr \hline 
6 & \left\{0,\frac{1}{2},\frac{9}{16},
\frac{1}{4},\frac{3}{4},
\frac{5}{16}  \right\}& 6_{\frac{7}{2},8.}^{16,656} &  177.(\text{factors}) \cr \hline 
6 & \left\{0,\frac{1}{2},
\frac{9}{16},\frac{1}{4},\frac{3}{4},\frac{13}{16}  \right\}& 6_{\frac{11}{2},8.}^{16,861} &  178.(\text{factors}) \cr \hline 
6 & \left\{0,
\frac{1}{2},\frac{15}{16},\frac{1}{4},\frac{3}{4},\frac{3}{16}  \right\}& 6_{\frac{1}{2},8.}^{16,818} &  179.(\text{factors}) \cr \hline 
6 & \left\{  0,\frac{1}{2},\frac{15}{16},\frac{1}{4},\frac{3}{4},
\frac{11}{16} \right\}& 6_{\frac{13}{2},8.}^{16,199} &  180.(\text{factors}) \cr \hline 
6 & \left\{  0,\frac{1}{2},\frac{3}{16},\frac{1}{4},
\frac{3}{4},\frac{7}{16}  \right\}& 6_{\frac{5}{2},8.}^{16,595} &  181.(\text{factors}) \cr \hline 
6 & \left\{ 0,\frac{1}{2},\frac{3}{16},\frac{1}{4},
\frac{3}{4},\frac{15}{16}  \right\}& 6_{\frac{1}{2},8.}^{16,156} &  182.(\text{factors}) \cr \hline 
6 & \left\{  0,\frac{1}{2},
\frac{5}{16},\frac{1}{4},\frac{3}{4},\frac{1}{16} \right\}& 6_{\frac{3}{2},8.}^{16,242} &  183.(\text{factors}) \cr \hline 
6 & \left\{  0,\frac{1}{2},\frac{5}{16},\frac{1}{4},
\frac{3}{4},\frac{9}{16} \right\}& 6_{\frac{7}{2},8.}^{16,684} &  184.(\text{factors}) \cr \hline 
6 & \left\{ 0,\frac{1}{2},\frac{11}{16},\frac{1}{4},
\frac{3}{4},\frac{7}{16}  \right\}& 6_{\frac{9}{2},8.}^{16,227} &  185.(\text{factors}) \cr \hline 
6 & \left\{  0,\frac{1}{2},\frac{11}{16},
\frac{1}{4},\frac{3}{4},\frac{15}{16}  \right\}&  6_{\frac{13}{2},8.}^{16,949}&  186.(\text{factors}) \cr \hline 
6 & \left\{  0,\frac{1}{2},
\frac{13}{16},\frac{1}{4},\frac{3}{4},\frac{1}{16} \right\}& 6_{\frac{15}{2},8.}^{16,553} &  187.(\text{factors}) \cr \hline 
6 & \left\{ 0,\frac{1}{2},\frac{13}{16},\frac{1}{4},
\frac{3}{4},\frac{9}{16}  \right\}& 6_{\frac{11}{2},8.}^{16,111} &  188.(\text{factors}) \cr \hline 
6 & \left\{ 0,\frac{1}{3},\frac{1}{3},\frac{1}{4},
\frac{7}{12},\frac{7}{12}   \right\}&  6_{3,6.}^{12,520}&  189.(\text{factors}) \cr \hline 
6 & \left\{  0,\frac{1}{3},\frac{1}{3},\frac{3}{4},
\frac{1}{12},\frac{1}{12} \right\}& 6_{1,6.}^{12,354} &  190.(\text{factors}) \cr \hline 
6 & \left\{ 0,\frac{2}{3},\frac{2}{3},\frac{1}{4},
\frac{11}{12},\frac{11}{12}  \right\}& 6_{7,6.}^{12,854} &  191.(\text{factors}) \cr \hline 
6 & \left\{  0,\frac{2}{3},
\frac{2}{3},\frac{3}{4},
\frac{5}{12},\frac{5}{12} \right\}& 6_{5,6.}^{12,208} &  192.(\text{factors}) \cr \hline 
\hline
\end{array}
$$
\caption{MTCs from 3-manifolds: Rank 6. Missing factored models.  }
\end{table}


\end{document}